\documentclass[11pt,canadian]{article}
\usepackage[letterpaper]{geometry}
\usepackage{color}
\usepackage{authblk}
\usepackage{xargs}[2008/03/08]
\usepackage{amsmath,amsthm,amsfonts}
\usepackage{graphicx}
\usepackage{subfig}
\usepackage{booktabs}
\usepackage{multirow}
\usepackage{ragged2e}
\usepackage{rotating}
\usepackage{array}
\usepackage{bm}
\usepackage{fancyhdr}

\usepackage[authoryear]{natbib}
\usepackage{xargs}[2008/03/08]
\usepackage{setspace}
\definecolor{note_fontcolor}{rgb}{0.800781, 0.800781, 0.800781}
\definecolor{pigment}{rgb}{0.2, 0.2, 0.6}

\usepackage{hyperref}
\hypersetup{unicode=true,
	bookmarksnumbered=false,bookmarksopen=false,
	breaklinks=false,pdfborder={0 0 0},pdfborderstyle={}, colorlinks=true,
	linkcolor=pigment, citecolor=pigment, urlcolor=pigment}

\makeatletter

\usepackage{color}
\definecolor{note_fontcolor}{rgb}{0.800781, 0.800781, 0.800781}

\definecolor{pigment}{rgb}{0.2, 0.2, 0.6}
\usepackage{arydshln}
\usepackage{breakcites}
\usepackage{breakurl}
\usepackage{amsfonts}
\usepackage{booktabs}

\@ifundefined{showcaptionsetup}{}{%
	\PassOptionsToPackage{caption=false}{subfig}}
\usepackage{subfig}
\makeatother

\global\long\def\argparentheses#1{\mathopen{\left(#1\right)}\mathclose{}}%
\global\long\def\ap#1{\argparentheses{#1}\mathclose{}}%

\global\long\def\smo#1{\sum_{#1}}%
\global\long\def\pro#1{\prod_{#1}}%

\global\long\def\mathtext#1{\mathrm{#1}}%
\global\long\def\mt#1{\mathtext{#1}}%

\global\long\def\argparentheses#1{\mathopen{\left(#1\right)}\mathclose{}}%
\global\long\def\mathtext#1{\mathrm{#1}}%
\global\long\def\mt#1{\mathtext{#1}}%

\newcommandx\der[3][usedefault, addprefix=\global, 1=, 2=, 3=]{\frac{d^{#2}#3}{d#1^{#2}}}%
\newcommandx\pder[3][usedefault, addprefix=\global, 1=, 2=]{\frac{\partial^{#2}#3}{\partial#1^{#2}}}%
\global\long\def\descr#1#2{\underset{#2}{\underbrace{#1}}}%
\global\long\def\e#1{{\scriptstyle \cdot10^{#1}}}%

\global\long\def\gammadist#1#2{\mt{Gamma}\ap{#1,\,#2}}%

\global\long\def\psame{\xi_{\mt{same}}}%
\global\long\def\pregion{\xi_{\mt{region}}}%
\global\long\def\pall{\xi_{\mt{all}}}%
\global\long\def\papp{\nu_{\mt{app}}}%
\global\long\def\preport{\nu_{\mt{record}}}%
\global\long\def\mmod{\mt{\ mod\ }}%

\author[1,2,*]{Samuel M. Fischer}
\author[3]{Pouria Ramazi}
\author[4]{Sean Simmons}
\author[5]{Mark S. Poesch}
\author[1,6]{Mark A. Lewis}

\affil[1]{Department of Mathematical and Statistical Sciences, University of Alberta, Edmonton, AB, T6G 2G1, Canada.}
\affil[2]{Department of Ecological Modelling, Helmholtz-Centre for Environmental Research - UFZ, Permoserstraße 15, 04318 Leipzig, Germany.}
\affil[3]{Department of Mathematics and Statistics, Brock University, St. Catharines, ON, L2S 3A1, Canada.}
\affil[4]{Angler’s Atlas, Goldstream Publishing, PO Box 182, Prince George, BC, V2L 4S1, Canada.}
\affil[5]{Department of Renewable Resources, University of Alberta, Edmonton, AB, T6G 2R3, Canada.}
\affil[6]{Department of Biological Sciences, University of Alberta, Edmonton, AB, T6G 2E9, Canada.}
\affil[*]{samuel.fischer@ualberta.ca}

\date{}

\begin{document}
	
	\chead{Boosting propagule transport models with app data}
	
	\title{Boosting Propagule Transport Models with Individual-Specific Data
		from Mobile Apps}
	\maketitle
	\begin{abstract}
		~
		\begin{enumerate}
			\item Management of invasive species and pathogens requires information
			about the traffic of potential vectors. Such information is often
			taken from vector traffic models fitted to survey data.  Here, user-specific
			data collected via mobile apps offer new opportunities to obtain more
			accurate estimates and to analyze how vectors' individual preferences
			affect propagule flows. However, data voluntarily reported via apps
			may lack some trip records, adding a significant layer of uncertainty.
			We show how the benefits of app-based data can be exploited despite
			this drawback.
			\item Based on data collected via an angler app, we built a stochastic
			model for angler traffic in the Canadian province Alberta. There,
			anglers facilitate the spread of whirling disease, a parasite-induced
			fish disease. The model is temporally and spatially explicit and accounts
			for individual preferences and repeating behaviour of anglers, helping
			to address the problem of missing trip records.
			\item We obtained estimates of angler traffic between all subbasins in
			Alberta. The model's accuracy exceeds that of direct empirical estimates
			even when fewer data were used to fit the model. The results indicate
			that anglers' local preferences and their tendency to revisit previous
			destinations reduce the number of long inter-waterbody trips potentially
			dispersing whirling disease. According to our model, anglers revisit
			their previous destination in $64\%$ of their trips, making these
			trips irrelevant for the spread of whirling disease. Furthermore,
			$54\%$ of fishing trips end in individual-specific spatially contained
			areas with mean radius of $54.7\,\text{km}$. Finally, although the
			fraction of trips that anglers report was unknown, we were able to
			estimate the total yearly number of fishing trips in Alberta, matching
			an independent empirical estimate.
			\item \textbf{Policy implications}: We make two major contributions: (1) we provide a model that uses
			mobile app data to boost the mechanistic accuracy of classic propagule
			transport models, and (2) we demonstrate the importance of individual-specific
			behaviour of vectors for propagule transport. Ignoring vectors' local
			preferences and their tendency to revisit previous destinations can
			lead to significant overestimates of vector traffic and biased estimates
			of propagule flows. This has clear implications for the management
			of invasive species and animal diseases. 
		\end{enumerate}
	\end{abstract}

	
	\begin{description}
		\item [{Keywords:}] Angler; Gravity Model; Invasives (Applied Ecology);
		Modelling (Disease Ecology); Smartphone Apps; Survey Method; Vector;
		Whirling Disease.
	\end{description}

	\section{Introduction}
	
	Recreational overland traffic is a major vector for several invasive
	species and pathogens \citep{karesh_wildlife_2005,hulme_trade_2009}.
	Examples include invasive plants and pathogens carried via the soil
	attached to gear and vehicles of tourists \citep{von_der_lippe_long-distance_2007,cushman_multi-scale_2008},
	invasive insects introduced along with campers' firewood \citep{koch_dispersal_2012},
	or invasive mussels, non-indigenous bait fish, and water-borne diseases
	spread by recreational boaters \citep{johnson_overland_2001} and
	anglers  \citep{nalepa2013quagga,litvak1993ecology,kilian2012assessment,gates2007myxospore}.
	Given the difficulties and costs associated with eradicating invasive
	species and pathogens once they have established at a site,  it
	is key that any management strategy prevents propagule  transport
	and detects new infestations early \citep{leung2007risk,pluess2012eradication}.
	This requires a detailed understanding of transport pathways and vector's
	movement patterns. 
	
	Data collected via smartphone apps have become a valuable resource
	to study human mobility \citep{wang2019extracting} and offer new
	opportunities to understand and predict the dispersal of invasive
	species and pathogens \citep{papenfuss_smartphones_2015,venturelli_angler_2017}.
	Assuming a sufficiently large user base, mobile app data can be
	collected at relatively low cost over large spatial and temporal scales
	\citep{papenfuss_smartphones_2015,venturelli_angler_2017}.
	However, even if many trip records are available, the datasets collected
	via apps are typically far from complete: often, only a small fraction
	of the population of interest, e.g. hikers or anglers, use any particular
	app, and app users do not record all their trips \citep{papenfuss_smartphones_2015}.
	Even if an app records thousands of trips, this number remains small
	in comparison to the vast number of origin-destination pairs for which
	traffic estimates may be desired. For example, if we seek to estimate
	the vector traffic between $100$ origins and $100$ destinations,
	the number of origin-destination \emph{pairs} is $10\,000$. Hence,
	direct empirical estimates of traffic flows can be prone to significant
	statistical error. 
	
	A common approach to bridge such data gaps is to use models, such
	as gravity models \citep{bossenbroek_prediction_2001,ferrari_gravity_2006,potapov_stochastic_2010,muirhead_evaluation_2011,li_validation_2011}.
	By combining empirical observations with additional covariates, e.g.
	geographical and socio-economic data, models can provide detailed
	estimates of vector traffic on broad scales and may even allow insights
	into the mechanisms behind traffic patterns. In the past, vector traffic
	models have been fitted to data collected via mail-out surveys \citep{potapov_stochastic_2010,muirhead_evaluation_2011,chivers_predicting_2012,drake_bycatch_2014},
	roadside traffic surveys \citep{fischer_hybrid_2020}, on-site surveys
	at origins and destinations \citep{leung_predicting_2004,bossenbroek_forecasting_2007},
	or registration records from origins and destinations \citep{bossenbroek_prediction_2001,prasad_modeling_2010}.
	However, since gathering data via these methods is often costly and
	the data may represent specific locations and time frames only, app
	data are a promising alternative resource for fitting vector traffic
	models. In this study, we show how this can be done.
	
	A drawback of app data is that app users may report their trips sparsely,
	making the temporal sequence of their trips incomplete.  The data
	may still yield insight into how often each destination is visited,
	but without knowing the full trip sequence, it is difficult to gauge
	how far and quickly vectors will spread propagules after being infested.
	A vector frequently revisiting their previous destination has a much
	lower risk of spreading a disease than a vector who prefers to alternate
	between destinations. As the risk of a successful transmission is
	highest between consecutively visited sites, disregarding the trips
	not recorded by app users could bias the predictions of propagule
	dispersion models significantly. 
	
	To estimate the number of relevant trips between sites despite missing
	data, some studies assume that the destinations of consecutive trips
	are chosen independently from one another \citep{bossenbroek_prediction_2001,leung_predicting_2004}.
	Then, incomplete trip sequences would be representative for all trips.
	In practice, however, individual travellers may have local preferences
	and tend to revisit previous destinations, which would lower the risk
	of propagule transport. Accounting for these individual preferences
	requires a more intricate modelling approach. We tackle this problem
	and build a vector movement model that can be fitted to incomplete
	app-based data. 
	
	Though our approach is applicable to studying the spread of various
	pests in both terrestrial and aquatic systems, we introduce and demonstrate
	it by considering a particular case study: we model angler movement
	in the Canadian province Alberta based on data collected via the ``MyCatch''
	angler app so as to create a risk map for the spread of whirling disease
	in Alberta. Our focal invader, whirling disease, is a fish disease
	caused by the aquatic parasite \textit{M. cerebralis} \citep{hofer1903uber},
	which can increase the death rate of juvenile salmonid fish up to
	$90\%$ \citep{elwell2009whirling} and may thus entail severe ecological
	and economic consequences \citep{turner2014whirling,ramazi_m_2021}.
	As there is currently no known cure for whirling disease in natural
	ecosystems \citep{turner2014whirling}, management is limited to reducing
	the risk of parasite introduction. 
	
	Our main objective is to estimate how often each subbasin in Alberta
	is visited by anglers who have visited an infested area on their previous
	trip. Whilst estimating angler traffic, we also assess how traffic
	depends on local preferences of individual anglers and their tendency
	to revisit previous destinations. Our results indicate that if these
	factors are not accounted for, local traffic is significantly underestimated
	while long-distance traffic is overestimated. This, in turn, has general
	implications for risk assessment and management of invasive species
	and infectious diseases.
	
	
	\section{Materials and Methods}
	
\begin{figure}
		\begin{centering}
			\includegraphics[scale=0.8]{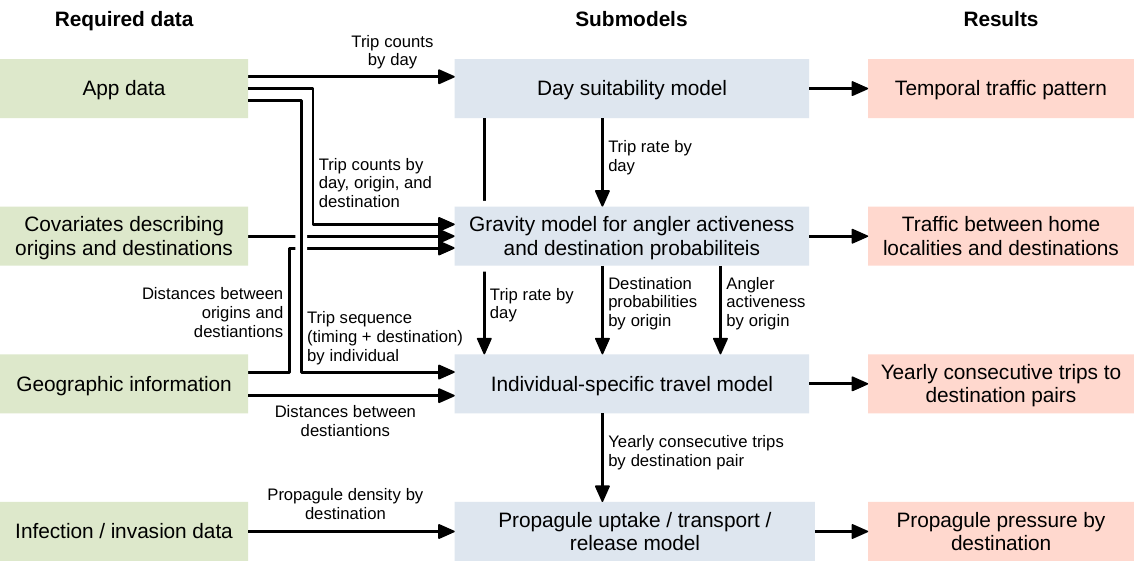}
			\par\end{centering}
		\caption{Model components. The data required for the analysis are displayed
			in green, the different submodels in blue, and the results in red.
			The submodels are combined into a stochastic traffic model described
			in section \ref{subsec:Angler-traffic-stochastic-process}. Though
			incorporating a sophisticated propagule transport model is possible,
			we use the number of directly consecutive trips from infested to uninfested
			areas as a proxy for propagule pressure in this study. \label{fig:components}}
	\end{figure}

	An overview of the data and submodels used in our approach is displayed
	in Figure \ref{fig:components}. Below we describe the components
	and their interplay in detail. An overview of the mathematical symbols
	used in this paper is provided in Appendix~S1 in Supporting Information. 
	
	\subsection{Data}
	
	We used a dataset collected via the MyCatch angler app, which can
	be downloaded free of charge for Android and iOS devices and  allows
	anglers to share information regarding waterbodies they visit, e.g.
	their catch success. App users need to provide their home postal codes
	and may record their fishing destinations either via GPS or select
	their destination waterbodies on a map. In addition to using the app,
	registered users can also enter information via a web interface. Though
	not all anglers in Alberta use the app, the app users have been found
	to be mostly representative of the province's anglers, with a slight
	bias towards higher app usage in urban areas \citep{johnston_comparative_2021}.
	
	The data were collected from May $2018$ to April $2020$ inclusive.
	We determined the home \emph{locality} (city, town, village, etc.)
	of each app user who recorded at least one trip within this time frame
	and collected the sequence of their fishing destinations along with
	the trip dates. If an angler recorded several trips to the same waterbody
	on a day, we merged these into a single trip. To keep the number of
	fishing destinations tractable, we aggregated them over \emph{subbasins}
	(hydrologic units of level $8$) and neglected more detailed information.
	Subbasins are a natural unit for modelling the spread of aquatic diseases,
	because they have a unique outflow each. Alberta consists of $422$
	subbasins with a mean area of $1517\text{ km}^{2}$. Our dataset included
	$575$ anglers, who made $2104$ trips. For $229$ of these trips,
	we could not determine the destination subbasin, because the anglers
	did not provide destination coordinates and the reported destination
	waterbodies spanned multiple subbasins. We disregarded these trips.
	All research was conducted in accordance with the Human Research Ethics
	Policy of the University of Alberta (approval number Pro00102610).
	
	As predictors for anglers' behaviour, we used data on the localities
	and the subbasins (Table~\ref{tab:Covariates-and-groups}). Besides
	geographical and socioeconomic data, we compiled data collected on
	the  Angler's Atlas website (\href{https://www.anglersatlas.com}{www.anglersatlas.com}).
	The website contains a page for each major waterbody in Alberta, providing
	anglers with waterbody-specific information and allowing them to report
	the species of fish they have caught there. Fish species reports can
	be upvoted and downvoted by other anglers to confirm or rebut an observation.
	We computed for each subbasin the area and perimeter of all waterbodies
	with at least one confirmed fish species. Furthermore, we computed
	the cumulative number of waterbody webpage visits and species upvotes
	per subbasin. For waterbodies spanning over multiple subbasins, we
	distributed the values over all applicable units according to their
	share of the waterbodies' perimeters. In addition to the listed covariates,
	we determined the number of registered anglers for each locality.
	The sources of the individual datasets we used are listed in the Data
	Sources section. 
	\begin{center}
		\begin{table}
			\begin{centering}
				\begin{tabular}{>{\raggedright}m{0.22\textwidth}>{\raggedright}p{0.45\textwidth}>{\centering}p{0.11\textwidth}>{\centering}p{0.11\textwidth}}
					\toprule 
					\textbf{Group} & \textbf{Covariate} & \textbf{Median} & \textbf{Maximum}\tabularnewline
					\midrule
					\midrule 
					\multirow{3}{0.22\textwidth}{Angler activeness in localities} & Locality population (2019) & $0.34\cdot10^{3}$ & $1\,286\cdot10^{3}$\tabularnewline
					& Mean income (2013) & $36\,900\,\text{\text{CAD}}$ & $99\,600\,\text{CAD}$\tabularnewline
					& Median income (2013) & $34\,600\,\text{CAD}$ & $78\,100\,\text{CAD}$\tabularnewline
					\midrule
					\multirow{4}{0.22\textwidth}{Fishing opportunities in subbasins} & Total perimeter of waterbodies & $350\,\text{km}$ & $1\,570\,\text{km}$\tabularnewline
					& Total area of waterbodies & $4\,\text{km}^{2}$ & $2\,160\,\text{km}^{2}$\tabularnewline
					& Total perimeter with confirmed species & $0\,\text{km}$ & $620\,\text{km}$\tabularnewline
					& Total area with confirmed species & $0\,\text{km}^{2}$ & $1\,390\,\text{km}^{2}$\tabularnewline
					\midrule
					\multirow{2}{0.22\textwidth}{Infrastructure in subbasins} & Population in $10\text{ km}$ range (2019) & $0.4\cdot10^{3}$ & $1\,394\cdot10^{3}$\tabularnewline
					& Public campgrounds in $10\text{ km}$ range & $1$ & $19$\tabularnewline
					\midrule
					\multirow{2}{0.22\textwidth}{Social media presence of subbasins} & Total species upvotes (2018 -- 2019) & $0$ & $196$\tabularnewline
					& Total waterbody web page visits (2018 -- 2019) & $140$ & $21\,945$\tabularnewline
					\bottomrule
				\end{tabular}
				\par\end{centering}
			\caption{Considered covariates and their median and maximum values. \label{tab:Covariates-and-groups}}
		\end{table}
		\par\end{center}

	\subsection{Angler traffic as a stochastic process\label{subsec:Angler-traffic-stochastic-process}}
	
\begin{figure}
		\begin{centering}
			\includegraphics[scale=0.8]{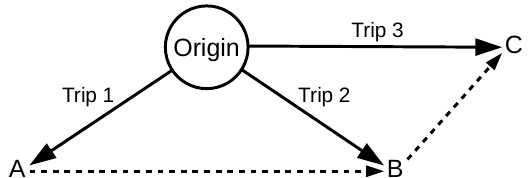}
			\par\end{centering}
		\caption{Possible sequence of trips to the destinations $A$, $B$, and $C$
			for an angler with home locality \textquotedblleft origin\textquotedblright .
			The risk that the angler transports propagules or pathogens is highest
			for consecutively visited fishing locations, i.e. for destinations
			$A$ and $B$ and for $B$ and $C$. Our goal is to estimate the number
			of such consecutive fishing trips for any pair of destinations. Our
			dataset contains information on individual trips, but some trips may
			not have been recorded. \label{fig:relevant-trips}}
	\end{figure}

	We modelled anglers' decision-making as a stochastic process, which
	determines both the \emph{recorded} number of fishing trips between
	each home locality and subbasin (\emph{origin} and \emph{destination})
	and the \emph{expected} yearly number $\mu_{j_{1}j_{2}}$ of trips
	anglers make to destination $j_{1}$ directly after visiting destination
	$j_{2}$. While we sought to estimate the latter number for all destination
	pairs, we fitted the model based on the former (Figure~\ref{fig:relevant-trips}). 
	
	We assumed that anglers start all their fishing trips at their home
	localities and visit a single destination per trip. They make trips
	randomly at rates dependent on their origins, the date, and random
	factors not explicitly covered in the model, e.g. weather conditions.
	We modelled the trip rate for an angler from origin $i$ on day $t$
	as $\mu_{i}\varepsilon_{t},$ where the \emph{angler activeness} $\mu_{i}$
	is the mean number of trips per day for an angler from origin $i$,
	and the \emph{day suitability }$\varepsilon_{t}$ is a gamma random
	variable with mean $\tau_{t}$, denoting how well day $t$ of the
	study period is suited for going fishing: 
	\begin{equation}
		\varepsilon_{t}\sim\gammadist{\frac{\tau_{t}}{\alpha}}{\alpha}.
	\end{equation}
	
	The gamma distribution can take on a variety of shapes and is thus
	suited for diverse modelling applications \citep{kleiber_statistical_2003,husak_use_2007}.
	The dispersion parameter $\alpha$ determines the variance of the
	day suitability; the expected day suitability $\tau_{t}$ is normalized
	so that its temporal average is $1$, i.e. $\frac{1}{T}\sum_{t}\tau_{t}=1$
	with $T$ being the number of days in the study period. We supposed
	that the day suitability $\varepsilon_{t}$ is the same for all anglers
	in Alberta, whereas their individual decisions are independent from
	one another. The values $\mu_{i}$ and $\tau_{t}$ are given by the
	submodels in section \ref{subsec:Submodels}. 
	
\begin{figure}
		\begin{centering}
			\includegraphics[scale=0.8]{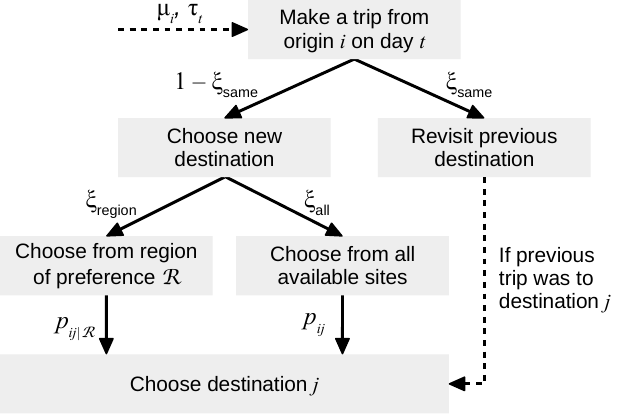}
			\par\end{centering}
		\caption{Visualization of anglers' decision-making process. The parameters
			$\mu_{i}$ and $\tau_{t}$ determine the expected rate at which anglers
			from origin $i$ make trips on day $t$. When an angler chooses their
			destination, they may revisit their previous destination with probability
			$\protect\psame$. Otherwise, they may either constrain their choice
			to their region of preference (with probability $\protect\pregion$)
			or make an unconstrained selection from all available destinations
			(with probability $\protect\pall$). If they decide to constrain their
			choice to their region of preference $\mathcal{R}$, they choose destination
			$j$ with probability $p_{ij|\mathcal{R}}$. Otherwise, they choose
			it with probability $p_{ij}$. \label{fig:choice-process}}
	\end{figure}

\begin{figure}
		\begin{centering}
			\includegraphics[scale=0.8]{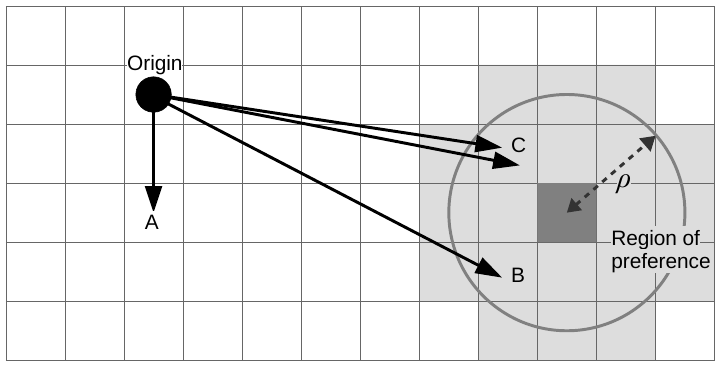}
			\par\end{centering}
		\caption{An example for a series of trips to destinations $A$, $B$, $C$,
			and again $C$ in order (depicted as black arrows) for an angler with
			the region of preference drawn in grey. Each grid cell represents
			a destination. The region of preference contains all destinations
			intersecting with the buffer of radius $\rho$ drawn around the centre
			of the destination coloured dark grey. The angler may choose any of
			the available destinations but often selects destinations within their
			region of preference. Furthermore, anglers may tend to revisit destinations
			on consecutive trips (e.g. destination C). Note that subbasins are
			not square grid cells in practice but can take any shape. \label{fig:trip-series}}
	\end{figure}

	We assumed that anglers choose the destinations of their trips based
	on individual local preferences and their previous fishing destinations
	(Figures~\ref{fig:choice-process}, \ref{fig:trip-series}). Consider
	an angler from origin $i$.  With probability $\psame$, they decide
	to revisit the destination of their last trip. Otherwise, they choose
	a new destination as follows: with probability $\pregion$, they constrain
	their destination choice to their \emph{region of preference} $\mathcal{R}$
	-- a spatially contained set of destinations that they personally
	like best -- and choose a destination $j\in\mathcal{R}$ according
	to probabilities $p_{ij|\mathcal{R}}$. Alternatively, with probability
	$\pall=1-\pregion$, they make an unconstrained choice from all available
	destinations $j$ according to probabilities $p_{ij}$.
	
	We supposed that each region of preference consists of destinations
	intersecting with a buffer of radius $\rho$ around a subbasin centre
	(Figure~\ref{fig:trip-series}).  Each angler's region of preference
	is fixed over time and chosen randomly. The probability $p_{i\mathcal{R}}$
	that an angler from origin $i$ has region of preference $\mathcal{R}$
	is proportional to how likely they would choose a destination in $\mathcal{R}$
	under the unconstrained strategy: 
	\begin{equation}
		p_{i\mathcal{R}}=\frac{1}{\smo{\tilde{\mathcal{R}}\in\mathfrak{R}}\smo{j\in\tilde{\mathcal{R}}}p_{ij}}\smo{j\in\mathcal{R}}p_{ij}.
	\end{equation}
	Here, $\mathfrak{R}$ is the set of all potential regions of preference.
	The probabilities $p_{ij|\mathcal{R}}$ are defined accordingly as
	\begin{equation}
		p_{ij|\mathcal{R}}=\frac{p_{ij}}{\smo{\tilde{j}\in\mathcal{R}}p_{i\tilde{j}}}.
	\end{equation}

	The assumptions above lead to a simplified model on an aggregate level.
	A random angler from origin $i$ will choose destination $j$ with
	probability 
	\begin{equation}
		\pregion\sum_{\mathcal{R}\in\mathfrak{R}}p_{i\mathcal{R}}p_{ij|\mathcal{R}}+\pall p_{ij}=p_{ij}
	\end{equation}
	unless they revisit their previous destination. Note that they cannot
	revisit a destination on their first trip. The probability that they
	choose destination $j$ on their second trip is therefore $(1-\psame)p_{ij}+\psame p_{ij}=p_{ij}$.
	By induction, the probability that a random angler from origin $i$
	chooses destination $j$ is $p_{ij}$. 
	
	Since not all anglers in Alberta used the MyCatch app and app users
	may not have recorded all their trips, an additional submodel for
	the sampling process is needed to incorporate the data recorded via
	the app. We assumed that each angler decided randomly to install and
	use the app with probability $\papp$, and that app users record a
	trip with probability $\preport$. 
	
	\subsection{Computing expected trip counts}
	
	Based on the model introduced above, the expected number of consecutive
	angler trips to destinations $j_{1}$ and $j_{2}$ can be computed
	as follows. Let $n_{i}$ be the number of anglers residing at origin
	$i$. Then, the expected number of trips that anglers from origin
	$i$ make during the study period is $n_{i}\mu_{i}T$. Now consider
	the probability that, for any pair of consecutive trips, $j_{1}$
	is the destination of the first trip and $j_{2}$ is the destination
	of the second trip. Recall that anglers may either revisit their previous
	location, constrain their destination choice to their region of preference,
	or choose their destination freely. Hence, the mean number of consecutive
	trips by anglers from origin $i$ to $j_{1}$ and $j_{2}$ during
	the study period is 
	\begin{multline}
		\mu_{ij_{1}j_{2}}=\descr{n_{i}\mu_{i}T}{\substack{\text{expected}\\
				\text{trip count}
			}
		}\Bigg(\hspace*{-0.2cm}\descr{\psame\delta_{j_{1}j_{2}}p_{ij_{1}}}{\substack{\text{prob. to travel \ensuremath{j_{1}\rightarrow j_{2}=j_{1}}}\\
				\text{by choosing \ensuremath{j_{1}} and revisiting it}\\
				\text{ without considering alternatives}
			}
		}\hspace*{-0.2cm}+\descr{\left(1-\psame\right)}{\substack{\text{prob. to consider}\\
				\text{alternatives to}\\
				\text{previous dest. \ensuremath{j_{1}}}
			}
		}\Bigg(\pall^{2}\hspace*{-0.2cm}\descr{p_{ij_{1}}p_{ij_{2}}}{\substack{\text{prob. to travel \ensuremath{j_{1}\rightarrow j_{2}}}\\
				\text{if both are chosen from}\\
				\text{all available sites}
			}
		}\\
		+\pall\pregion\Bigg(\hspace*{-0.4cm}\descr{p_{ij_{1}}\smo{\mathcal{R}\,:\,j_{2}\in\mathcal{R}}p_{i\mathcal{R}}p_{ij_{2}|\mathcal{R}}}{\substack{\text{prob. to travel \ensuremath{j_{1}\rightarrow j_{2}} if \ensuremath{j_{1}} is}\\
				\text{chosen from all available sites}\\
				\text{and \ensuremath{j_{2}} from a region of preference}
			}
		}\hspace*{-0.1cm}+\hspace*{-0.1cm}\descr{p_{ij_{2}}\smo{\mathcal{R}\,:\,j_{1}\in\mathcal{R}}p_{i\mathcal{R}}p_{ij_{1}|\mathcal{R}}}{\substack{\text{prob. to travel \ensuremath{j_{1}\rightarrow j_{2}} if \ensuremath{j_{1}} is}\\
				\text{chosen from a region of preference}\\
				\text{and \ensuremath{j_{2}} from all available sites}
			}
		}\hspace*{-0.4cm}\Bigg)+\pregion^{2}\hspace*{-0.2cm}\descr{\smo{\mathcal{R}\,:\,j_{1},j_{2}\in\mathcal{R}}p_{i\mathcal{R}}p_{ij_{1}|\mathcal{R}}p_{ij_{2}|\mathcal{R}}}{\substack{\text{prob. to travel \ensuremath{j_{1}\rightarrow j_{2}} if both are}\\
				\text{chosen from a region of preference}
			}
		}\hspace*{-0.1cm}\Bigg)\Bigg).\label{eq:mean_HUC_HUC_traffic}
	\end{multline}
	Here, $\delta_{j_{1}j_{2}}$ is $1$ if $j_{1}=j_{2}$ and $0$ otherwise.
	The right hand side of equation (\ref{eq:mean_HUC_HUC_traffic}) can
	be simplified to speed up computations, as we show in Appendix~S2. 
	
	We computed the expected number $\mu_{j_{1}j_{2}}$ of consecutive
	trips to destinations $j_{1}$ and $j_{2}$ by summing $\mu_{ij_{1}j_{2}}$
	over all origins $i$. 
	To determine how many anglers access a destination $j_{2}$ after
	having visited a whirling-disease infested site, we furthermore summed
	the $\mu_{ij_{1}j_{2}}$ over all subbasins $j_{1}$ where the disease
	is present already. Based on these results, we also computed the number
	of these trips for each origin $i$.
	
	\subsection{Submodels for day suitability, angler activeness, and destination
		probabilities\label{subsec:Submodels}}
	
	The expected day suitability $\tau_{t}$ may change in weekly and
	seasonal cycles. We modelled these variations using the probability
	density function $f_{\mt{vM}}\ap{\cdot;\kappa,\theta}$ of the von
	Mises distribution, which is a cyclic distribution resembling the
	normal distribution \citep{lee_circular_2010}. The shape of the function
	is controlled via the two parameters $\theta$, determining the location
	of the mode, and $\kappa$, determining how sharp the maximum is.
	We defined the expected suitability $\tau_{t}$ of day $t$ as follows:
	\begin{equation}
		\tau_{t}=c_{\mt{norm}}\descr{\left(c_{\mt{week}}+f_{\mt{vM}}\ap{2\pi\frac{t\mmod7}{7};\theta_{\mt{week}},\kappa_{\mt{week}}}\right)}{\text{weekly variations}}\descr{\left(c_{\mt{year}}+f_{\mt{vM}}\ap{2\pi\frac{t\mmod365}{365};\theta_{\mt{year}},\kappa_{\mt{year}}}\right)}{\text{yearly variations}}.\label{eq:expected-day-suitability}
	\end{equation}
	The constants $c_{\mt{year}}$ and $c_{\mt{week}}$ are parameters
	controlling the amplitude and vertical shift of the weekly and seasonal
	cycles; the constant $c_{\mt{norm}}$ is chosen so that $\frac{1}{T}\sum_{t}\tau_{t}=1$.
	If the study period includes leapyears, $\tau_{t}$ must be adjusted
	accordingly.
	
	As the $2104$ trips in our dataset did not suffice to estimate the
	choice probabilities $p_{ij}$ for all $180,000$ pairs of localities
	and subbasins directly, we estimated $\mu_{i}$ and $p_{ij}$ based
	on covariates on 
	the origins and destinations. To that end, we applied the framework
	of gravity models. Gravity models estimate the mean number of trips
	between each origin and destination as a product of (1) the \emph{repulsiveness}
	of the origin, proportional to the number of outbound trips; (2) the
	\emph{attractiveness} of the destination, proportional to the number
	of inbound trips; and (3) a decaying function of the distance between
	origin and destination. Repulsiveness and attractiveness are typically
	functions of covariates characterizing the origins and destinations.
	In our model, the expected outbound traffic of origin $i$ is given
	by the product $n_{i}\mu_{i}$ of angler count and activeness, and
	the expected traffic between origin $i$ and destination $j$ is $n_{i}\mu_{i}p_{ij}$.
	Therefore, the repulsiveness of origin $i$ corresponds to the product
	$n_{i}\mu_{i}$, whereas the product of distance decay function and
	attractiveness defines the choice probabilities $p_{ij}$. Let $d_{ij}$
	be the linear distance between origin $i$ and destination $j$, and
	let $a_{j}$ be the attractiveness of destination $j$, measuring
	both the quantity and quality of fishing opportunities. Then, 
	\begin{equation}
		p_{ij}=\frac{a_{j}D\ap{d_{ij}}}{\smo{\tilde{j}}a_{\tilde{j}}D\ap{d_{i\tilde{j}}}},\label{eq:city-HUC-mean}
	\end{equation}
	with the distance decay function $D$, which we define as 
	\begin{equation}
		D\ap{d_{ij}}=\frac{d_{0}^{\gamma_{\mt{distance}}}}{d_{0}^{\gamma_{\mt{distance}}}+d_{ij}^{\gamma_{\mt{distance}}}}.\label{eq:distance-decay}
	\end{equation}
	The parameter $d_{0}$ is the half saturation constant, given as the
	distance at which $D\ap{d_{ij}}=\frac{1}{2}$.
	
	To define an appropriate function to compute $\mu_{i}$ and $a_{j}$
	based on the covariates, we categorized the covariates into groups
	$\mathcal{X}$ (Table~\ref{tab:Covariates-and-groups}), each accounting
	for a different component that is necessary for high angler traffic
	between an origin and a destination \citep[cf.][]{fischer_hybrid_2020}.
	For each origin or destination $k$, we assigned the score $(\beta_{\bm{x}}x_{k})^{\gamma_{\bm{x}}}$
	to each covariate $\bm{x}\in\mathcal{X}$, where $x_{k}$ is the component
	of $\bm{x}$ corresponding to $k$ and the parameters $\beta_{\bm{x}}$
	and $\gamma_{\bm{x}}$ describe the impact of $\bm{x}$. We then added
	these individual scores to obtain a score for each group $\mathcal{X}$,
	so that a high score for \emph{one} covariate suffices to make the
	group's score large. Finally, we multiplied the scores for the different
	groups, making a high score for \emph{all} components necessary to
	boost the number of angler trips. With scaling constant $c$, we set
	\begin{align}
		\mu_{i} & =c\pro{\substack{\text{origin covariate}\\
				\text{groups }\mathcal{X}
			}
		}\left(1+\smo{\bm{x}\in\mathcal{X}}\left(\beta_{\bm{x}}x_{i}\right)^{\gamma_{\bm{x}}}\right),\label{eq:trip-rate}\\
		a_{j} & =\pro{\substack{\text{destination covariate}\\
				\text{groups }\mathcal{X}
			}
		}\left(1+\smo{\bm{x}\in\mathcal{X}}\left(\beta_{\bm{x}}x_{j}\right)^{\gamma_{\bm{x}}}\right),\label{eq:attractiveness}
	\end{align}
	where the \textit{origin} and \textit{destination covariate groups}
	are the locality and subbasin groups in Table~\ref{tab:Covariates-and-groups}.

	\subsection{Fitting the model}
	
	We fitted the model via maximizing the likelihood associated with
	the recorded app data. However, fitting the complete model all at
	once is computationally costly due to the complicated form of the
	likelihood function and the large number of parameters. Therefore,
	we eliminated parameters by summing over certain quantities to obtain
	submodels with simpler likelihood functions. Furthermore, we made
	approximations via independence assumptions, disregarding the identity
	of anglers in some fitting stages (see below). Since most trips are
	made by independent anglers, our parameter estimates remain valid
	despite these simplifications \citep{varin_composite_2008}. 
	
	We fitted the model in three steps: first, we considered the submodel
	for the day suitability $\varepsilon_{t}$; second, we estimated the
	angler activeness $\mu_{i}$ and the destination choice probabilities
	$p_{ij}$; and third, we estimated the parameters $\psame$, $\pregion$,
	$\pall$, $\papp$, $\preport$, and $\rho$ modelling anglers' tendencies
	to constrain their trip choices and to record trips. Below, we briefly
	explain each of these steps; more details can be found in Appendix~S3. In
	each of the steps, we exploited that (1) a Poisson random variable
	with a gamma distributed mean is negative binomially distributed and
	that (2) the mixture of a negative binomial and a binomial distribution
	remains negative binomially distributed \citep{villa_using_2006}. 
	
	\subsubsection{Day suitability}
	
	We estimated the expected day suitability $\tau_{t}$ by fitting the
	distribution of the total number $N_{t}$ of recorded angler trips
	on day $t$ to the data. According to our model, $N_{t}$ is negative
	binomially distributed with dispersion parameter $\frac{\alpha}{\tau_{t}}$
	and mean $\preport\tau_{t}\smo i\tilde{n}_{i}\mu_{i}$, where $\tilde{n}_{i}$
	is the number of app users in locality $i$. As $\tilde{n}_{i}$ is
	a random variable itself and constant over the study period, it is
	not straightforward to derive the exact distribution of $N_{t}$.
	However, since $N_{t}$ describes the aggregate trip counts of many
	anglers, who rarely make more than one trip per day, it is reasonable
	to consider trips as mutually independent on each day.  Then, the
	distribution of $N_{t}$ is negative binomially distributed with dispersion
	$\frac{\alpha}{\tau_{t}}$ and mean $\tau_{t}\bar{\mu}$, where $\bar{\mu}=\papp\preport\smo in_{i}\mu_{i}$.
	Hence, by fitting the distribution of $N_{t}$, we obtained estimates
	for the parameters $\alpha$, $\bar{\mu}$, and those controlling
	the shape of $\tau_{t}$. See Appendix~S3.1 for further details.

	\subsubsection{Angler activeness and destination choice probabilities }
	
	To estimate the angler activeness values $\mu_{i}$ and the destination
	choice probabilities $p_{ij}$, we considered the trip counts $N_{ijt}$
	for origin-destination pairs $\left(i,j\right)$ and days $t$. We
	fitted the joint distribution of the $N_{ijt}$ to our data for all
	origin-destination pairs and days of the study period. With the independence
	approximation from the previous section, each $N_{ijt}$ follows a
	negative binomial distribution with dispersion parameter $\frac{\alpha}{\tau_{t}}$
	and mean $\papp\preport\tau_{t}n_{i}\mu_{i}p_{ij}$. To improve the
	computational performance, we also considered the trips from different
	localities as mutually independent. The values $\tau_{t}$ were known
	from the previous fitting stage. By fitting $N_{ijt}$ to the observed
	values, we obtained estimates for the parameters of $\mu_{i}$ and
	$p_{ij}$. The scaling constant $c$ and the probabilities $\papp$
	and $\preport$ are not identifiable in this fitting stage, and we
	replaced them with a parameter $C=\papp\preport c$ here. Refer to
	Appendix~S3.2 for a method to compute the likelihood efficiently.

	\subsubsection{Remaining parameters}
	
	To fit the choice parameters $\psame$, $\pregion$, $\pall$, $\papp$,
	and $\preport$, we considered each angler and their trips individually.
	First, we determined the likelihood for the temporal sequence of
	their trips. Then we computed the likelihood for their destination
	choices given the timing of the trips. Because the destination choices
	for consecutive trips are not independent and we need to consider
	unknown numbers of intermediate unrecorded trips, the likelihood function
	has a complicated form involving convolutions and special functions.
	Nonetheless, it can be computed numerically with reasonable effort
	if partial intermediate results are reused where possible. We refer
	the reader to Appendix~S3.3 for details. Dependencies between trips of \emph{different}
	anglers were disregarded at this stage so as to facilitate efficient
	computation.
	
	To fit the radius $\rho$ of the inscribed circle of anglers' regions
	of preferences, we conducted a grid search with steps of $1\,\text{km}$
	in the interval between $10\,\text{km}$ and $80\,\text{km}$. For
	each considered value of $\rho$, we maximized the likelihood with
	respect to the remaining parameters; finally, we chose the radius
	leading to the maximal likelihood. We conducted a grid search, because
	the regions of preference are discrete entities, making gradient descent
	methods inapplicable to fit $\rho$.

	\subsubsection{Optimization methods and model selection}
	
	We used a combination of multiple optimization algorithms to maximize
	the likelihood. We applied the differential evolution algorithm \citep{storn_differential_1997}
	for a global search of the parameter space, and improved upon the
	results via gradient-based search algorithms \citep{kraft_software_1988,byrd_limited_1995,nocedal_trust-region_2006}.
	Details can be found in Appendix~S4. We implemented the model in the programming
	language \texttt{Python} (version 3.7) along with the \texttt{Scipy}
	libraries \citep{jones_scipy:_2001}. 
	
	To decide which covariates and parameters should be included in the
	model without overfitting, we used the information criterion by \citet{akaike_new_1974}
	(AIC). This metric is particularly suitable if the modelling goal
	is prediction \citep{ghosh_model_2001}.  When fitting the day suitability
	function $\tau_{t}$, we considered simplified models with the parameters
	$c_{\mt{week}}$, $c_{\mt{year}}$, $\kappa_{\mt{week}}$, and $\kappa_{\mt{year}}$,
	(equation (\ref{eq:expected-day-suitability})) set to $0$, respectively.
	For the angler activeness $\mu_{i}$ and destination choice probabilities
	$p_{ij}$, we considered models with any combination of the parameters
	$\beta_{\bm{x}}$ and $\gamma_{\bm{x}}$ (equations (\ref{eq:trip-rate})-(\ref{eq:attractiveness}))
	set to zero. Only for covariates $\bm{x}$ that had $0$-values for
	some origins or destinations, we tested models with $\gamma_{\bm{x}}=1$
	instead. We furthermore tested models without the parameter $d_{0}$
	(equation (\ref{eq:distance-decay})). We searched for the model with
	the minimal AIC value by using a branch and bound algorithm (Appendix~S4.2).
	This allowed us to find the optimal model without having to consider
	all potential candidates. 
	
	Note that we made approximations via independence assumptions, which
	violate the underlying assumptions of AIC. Hence, the metric may tend
	to favour overfitting models. We therefore chose the simplest model
	among those with AIC values less than $10$ units higher than the
	minimal AIC. Models whose AIC value is more than $10$ units higher
	than the minimal AIC may have little empirical support \citep{burnham_multimodel_2004}.

	\subsubsection{Model evaluation}
	
	We evaluated the trustworthiness of our parameter estimates by computing
	confidence intervals using a method based on the profile likelihood
	\citep{fischer_robust_2021}. Note that since we did not evaluate
	the joint model all at once and made approximations via independence
	assumptions, the true confidence intervals may be larger. Nonetheless,
	the approximate confidence intervals are suited to detect estimability
	issues and problems arising from multicollinearity of covariates. 
	
	To validate the submodel for angler traffic between localities and
	subbasins, we randomly split the app data into a training (fitting)
	and a testing (validation) dataset, each containing observations for
	half of the anglers, respectively. Note that a random split is in
	line with our purpose of evaluating the model accuracy in predicting
	unreported trips, and hence, a temporal split used for evaluating
	the model accuracy in making future predictions is not needed \citep{ramazi_predicting_2021}.
	We fitted our model to the training data and computed the mean yearly
	number of recorded angler trips for each origin, destination, and
	origin-destination pair. Then we plotted the predicted values against
	the observed values. 
	
	The purpose of the submodel for locality-to-subbasin traffic was to
	fill data gaps stemming from the limited number of angler trips in
	our dataset. To ensure that the model is suited to fill these gaps
	without introducing additional error, we computed the mean absolute
	errors between the submodel's results and the observations from the
	testing data. We then compared the resulting values with those obtained
	by using direct estimates from the training data without an additional
	model. 
	
	Our model yields absolute estimates of angler traffic based on voluntarily
	reported trip data without using a priori information on how many
	trips anglers actually made. To assess the accuracy of our estimates
	of the trip frequency, we compared the number of days anglers go fishing
	per year as per our model (see Appendix~S5) with empirical data collected
	in a mail-out survey by the Department of Fisheries and Oceans Canada
	in $2016$ \citep{dfo_survey_2019}. 
	
	To facilitate a qualitative comparison of our model's accuracy with
	predictions from similar studies, we computed Nagelkerke's pseudo-$R^{2}$
	\citep{nagelkerke_note_1991}. This measure indicates how well the
	model performs in comparison to a non-informative null model. In contrast
	to the classical $R^{2}$, Nagelkerke's pseudo-$R^{2}$ can be applied
	even if the data are not assumed to be normally distributed with identical
	standard deviations. We computed pseudo-$R^{2}$ values for each model
	stage: the day suitability model; the joint day suitability, angler
	activeness and destination choice model; and the complete model with
	all submodels. As null models, we used negative binomial distributions
	treating all days, localities, and subbasins similarly. The parameters
	$\psame$ and $\pregion$ capturing anglers' local preferences were
	set to zero.

	\section{Results}
	
\begin{figure}
		\begin{centering}
			\includegraphics[width=1\textwidth]{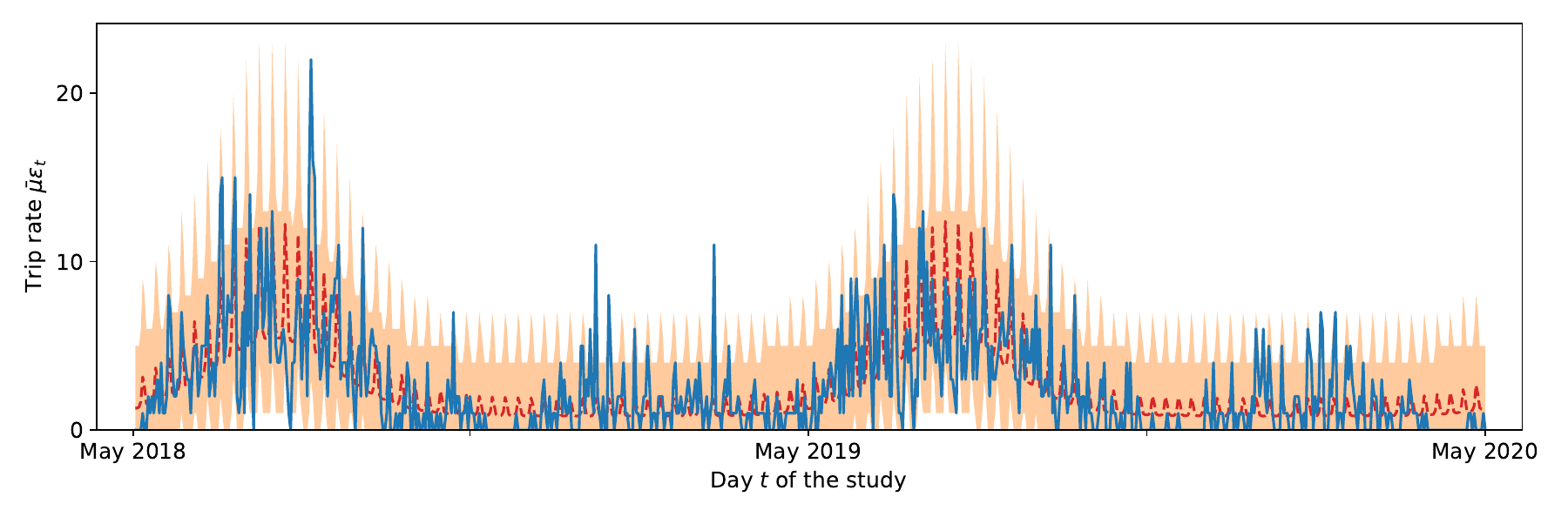}
			\par\end{centering}
		\caption{Observed and modelled trip rates for each day of the study period.
			The observed number of trips $N_{t}$ is drawn as solid blue line,
			the predicted mean of $N_{t}$ as dashed red line, and the predicted
			$95\%$ confidence range as light red area. \label{fig:day-trips}}
	\end{figure}

	The selected day suitability model contained all considered parameters
	(Table~\ref{tab:day-parameters}). The traffic was estimated highest
	on Saturdays and to peak on July $14$. The rates of fishing trips
	during the weekly peak was estimated to be $2.27$ times higher than
	on the weekly low; yearly cycles changed the expected traffic rate
	by up to factor $6.6$. The confidence intervals were relatively narrow
	for most of the parameters; only the shape parameter for the weekly
	cycles had a large upper confidence interval bound.  The temporal
	distribution of trips and the expected trip rates estimated by the
	model are displayed in Figure~\ref{fig:day-trips}. 
	
	\begin{table}[t]
		\begin{centering}
			\footnotesize \setlength\arrayrulewidth{0.5pt}
			\par\end{centering}
		\begin{centering}
			\begin{tabular}{>{\raggedright}p{1.2cm}>{\raggedright}p{9.2cm}>{\centering}p{2cm}>{\centering}p{1.2cm}>{\centering}p{1.2cm}}
				\toprule 
				\textbf{Para-meter} & \textbf{Explanation} & \textbf{Estimate} & \multicolumn{2}{>{\centering}p{2.4cm}}{\textbf{Confidence Interval}}\tabularnewline
				\midrule
				\midrule 
				$\alpha$ & Dispersion parameter & $0.33$ & $0.25$ & $0.42$\tabularnewline
				\midrule 
				$\bar{\mu}$ & Mean total recorded trips per day & $2.56$ & $1.52$ & $4.34$\tabularnewline
				\midrule 
				$c_{\mt{week}}$ & Week addition constant & $0.45$ & $0.24$ & $0.7$\tabularnewline
				$\theta_{\mt{week}}$ & Week location constant & $5.64$ & $5.5$ & $5.8$\tabularnewline
				$\kappa_{\mt{week}}$ & Week shape constant & $2.87$ & $1.36$ & $\infty$\tabularnewline
				\midrule 
				$c_{\mt{year}}$ & Year addition constant & $0.12$ & $0.1$ & $0.15$\tabularnewline
				$\theta_{\mt{year}}$ & Year location constant & $3.35$ & $3.29$ & $3.42$\tabularnewline
				$\kappa_{\mt{year}}$ & Year shape constant & $7.4$ & $6.53$ & $8.29$\tabularnewline
				\bottomrule
			\end{tabular}
			\par\end{centering}
		\caption{Parameters and estimates along with approximate $95\%$ confidence
			intervals after fitting the model to the daily trip counts. \label{tab:day-parameters}}
	\end{table}
	
	The angler activeness and destination choice model with the least
	number of parameters and $\Delta\mt{AIC}\leq10$ included $15$ parameters
	(Table~\ref{tab:gravity-parameters}; $\Delta\mt{AIC}=1.5$). The
	model with minimal AIC included the number of website visits as an
	additional covariate. The selected model uses localities' population
	counts and the mean income numbers to estimate angler activeness.
	The angler activeness varied by up to factor $4.5$ among localities.
	For a locality with median characteristics, a population increase
	of $1000$ increases angler activeness by $6\%$, and a mean income
	increase by $\text{CAD}\,1000$ increases activeness by $2\%$. 
	
	The localities from which anglers make the most consecutive trips
	between infested and uninfested subbasins were the population centres
	Calgary and Edmonton. Setting the trip count into relation with the
	number of registered anglers, Calgary, located in direct proximity
	to the infested area, had higher relevant traffic with $2.86$ high-risk
	trips per registered angler and year as compared to $0.65$ for Edmonton.
	Considering all inhabitants, rural municipalities had higher relative
	trip counts: the $100$ localities with the most high-risk trips per
	inhabitant had less than $8\,000$ inhabitants. 
	
	The subbasin attractiveness was estimated based on the perimeter of
	the waterbodies in the subbasins, the surface area of the waterbodies
	with confirmed species, the number of campgrounds, and the website
	species upvotes. The attractiveness values varied greatly among subbasins,
	by up to factor $1179$. For a subbasin with median characteristics,
	an increase in the total waterbody perimeter by $500\,\text{km}$
	increased attractiveness by $8.7\%$, whereas an increase of $1000\,\text{km}$
	increased attractiveness by factor $3$. If the total area of waterbodies
	with confirmed species were increased by $10\,\text{km}^{2}$, attractiveness
	would rise by $69\%$. An additional campground increased attractiveness
	by $20\%$, and an additional positive species vote by $180\%$. The
	traffic between localities and subbasins decreased in square order
	of their distance; i.e., a subbasin twice as far as a similar subbasin
	was only $25\%$ as likely to be chosen as fishing destination.
	\begin{table}[t]
		\begin{centering}
			\begin{tabular}{>{\raggedright}p{1.8cm}>{\raggedright}p{9cm}>{\centering}p{2cm}>{\centering}p{1.2cm}>{\centering}p{1.2cm}}
				\toprule 
				\textbf{Para-meter} & \textbf{Explanation} & \textbf{Estimate} & \multicolumn{2}{>{\centering}p{2.4cm}}{\textbf{Confidence Interval}}\tabularnewline
				\midrule
				\midrule 
				$\alpha$ & Dispersion parameter & $6.99$ & $4.91$ & $9.59$\tabularnewline
				\midrule 
				$C$ & Scaling constant for the mean daily number of recorded trips  & $1\e{-9}$ & $3.51\e{-11}$ & $5.12\e{-8}$\tabularnewline
				\midrule 
				$\delta_{0}$ & Distance of half choice-probability decay $[\text{km}]$ & $28.74$ & $22.02$ & $35.91$\tabularnewline
				$\gamma_{\mt{distance}}$ & Distance exponent & $2.09$ & $1.97$ & $2.22$\tabularnewline
				\midrule 
				$\beta_{\mt{population}}$ & City population factor $[1/10^{3}]$ & $6.95\e{29}$ & $4.29\e{29}$ & $1.23\e{30}$\tabularnewline
				$\gamma_{\mt{population}}$ & City population exponent & $0.14$ & $0.09$ & $0.16$\tabularnewline
				\midrule 
				$\beta_{\mt{mean\,income}}$ & Mean income factor $[1/(10^{3}\text{CAD})]$ & $909.1$ & $11.61$ & $2.3\e 6$\tabularnewline
				$\gamma_{\mt{mean\,income}}$ & Mean income exponent (not included in the model) & $1$ & -- & --\tabularnewline
				\midrule 
				$\beta_{\mt{perimeter}}$ & Water perimeter factor $[1/(10^{3}\text{km})]$ & $0.82$ & $0.76$ & $0.94$\tabularnewline
				$\gamma_{\mt{perimeter}}$ & Water perimeter exponent & $6.76$ & $3.89$ & $9.62$\tabularnewline
				\midrule 
				$\beta_{\mt{area\,confirmed}}$ & Water area confirmed factor $[1/(10^{3}\text{km}^{2})]$ & $38.95$ & $9.25$ & $305.8$\tabularnewline
				$\gamma_{\mt{area\,confirmed}}$ & Water area confirmed exponent  & $0.4$ & $0.24$ & $0.67$\tabularnewline
				\midrule 
				$\beta_{\mt{campground}}$ & Campground factor $[1]$ & $0.25$ & $0.12$ & $0.62$\tabularnewline
				$\gamma_{\mt{campground}}$ & Campground exponent  & $1$ & $0.65$ & $1.46$\tabularnewline
				\midrule 
				$\beta_{\mt{species\,vote}}$ & Species vote factor $[1]$ & $2.8$ & $1$ & $8.4$\tabularnewline
				$\gamma_{\mt{species\,vote}}$ & Species vote exponent  & $0.57$ & $0.5$ & $0.65$\tabularnewline
				\bottomrule
			\end{tabular}
			\par\end{centering}
		\caption{Parameters and estimates along with approximate $95\%$ confidence
			intervals after fitting the model for angler activeness and destination
			choice. Note that though we report all parameters on the original
			scale, we worked with log-transformed parameters internally to avoid
			numerical errors due to extreme parameter values. \label{tab:gravity-parameters}}
	\end{table}
	
	The estimates for the remaining choice parameter are displayed in
	Table~\ref{tab:choice-parameters}. The fraction of anglers using
	the app was estimated to be $0.22\%$, and the estimated probability
	that app users report a trip was $0.05$. The dispersion parameter,
	modelling the impact of stochastic events on anglers' daily trip rates
	was estimated $11.8$. The model predicts that on $64\%$ of their
	trips, anglers revisit their last destination. They choose a destination
	in their region of preference in $54\%$ of their trips and choose
	the destinations for the remaining $46\%$ trips from all over Alberta.
	The estimated radius $\rho$ for the inscribed circle of regions of
	preference was $31\,\text{km}$. This translates to a mean radius
	of $54.7\,\text{km}$ for the regions of preference. The parameter
	confidence intervals were relatively narrow except for the probability
	that app users report a trip (Table~\ref{tab:choice-parameters}).
	\begin{table}[t]
		\begin{centering}
			\begin{tabular}{>{\raggedright}p{1.2cm}>{\raggedright}p{9.2cm}>{\centering}p{2cm}>{\centering}p{1.2cm}>{\centering}p{1.2cm}}
				\toprule 
				\textbf{Para-meter} & \textbf{Explanation} & \textbf{Estimate} & \multicolumn{2}{>{\centering}p{2.4cm}}{\textbf{Confidence Interval}}\tabularnewline
				\midrule
				\midrule 
				$\alpha$ & Dispersion parameter &    $11.81$ & $8.65$ & $15.6$\tabularnewline
				\midrule 
				$\psame$ & Probability to revisit the previous destination &    $0.64$ & $0.53$ & $0.79$\tabularnewline
				\midrule 
				$\pregion$ & Probability to constrain the destination choice to the region of preference &   $0.54$ & $0.5$ & $0.59$\tabularnewline
				\midrule 
				$\papp$ & Probability to use the app & $0.0022$ & $0.0021$ & $0.0023$\tabularnewline
				\midrule 
				$\preport$ & Probability to record a trip &  $0.052$ & $0.022$ & $0.104$\tabularnewline
				\bottomrule
			\end{tabular}
			\par\end{centering}
		\caption{Parameters and estimates along with approximate $95\%$ confidence
			intervals after fitting the model for the individual angler choices
			\label{tab:choice-parameters}}
	\end{table}
	
\begin{figure}
		\subfloat[\label{fig:HUC-HUC-trips}]{\includegraphics[width=0.48\textwidth]{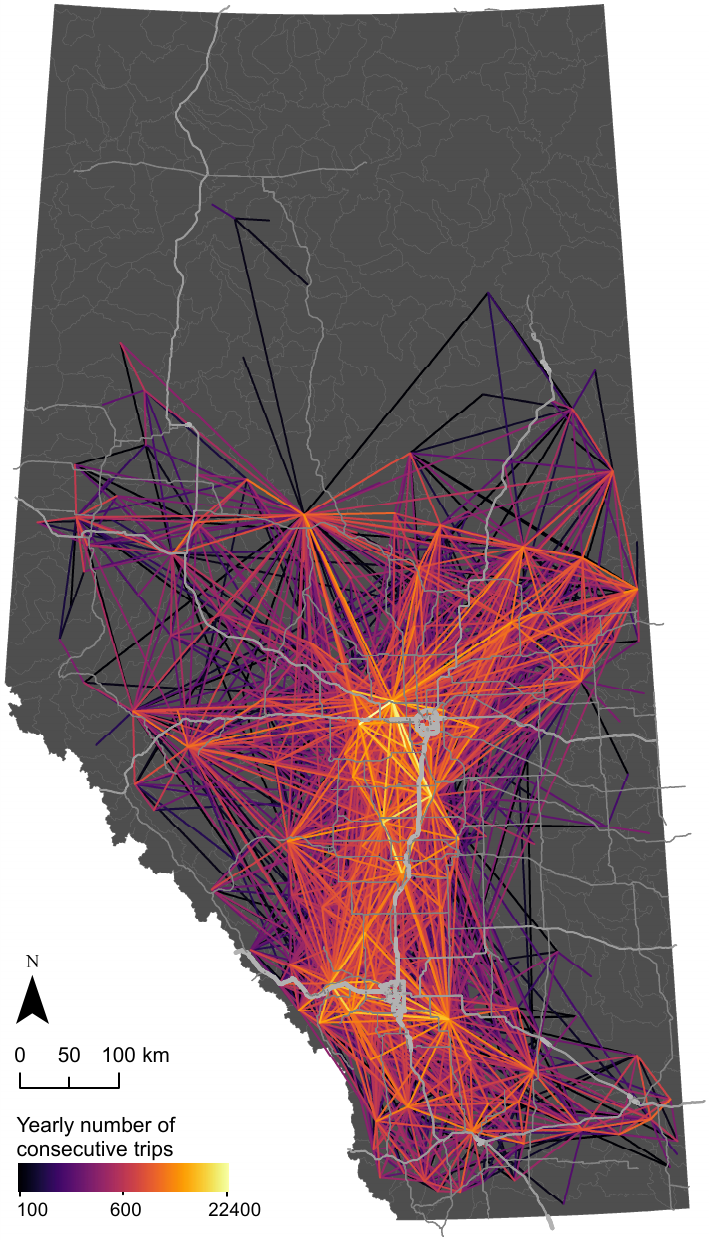}}\hspace*{\fill}\subfloat[\label{fig:HUC-risk}]{\includegraphics[width=0.48\textwidth]{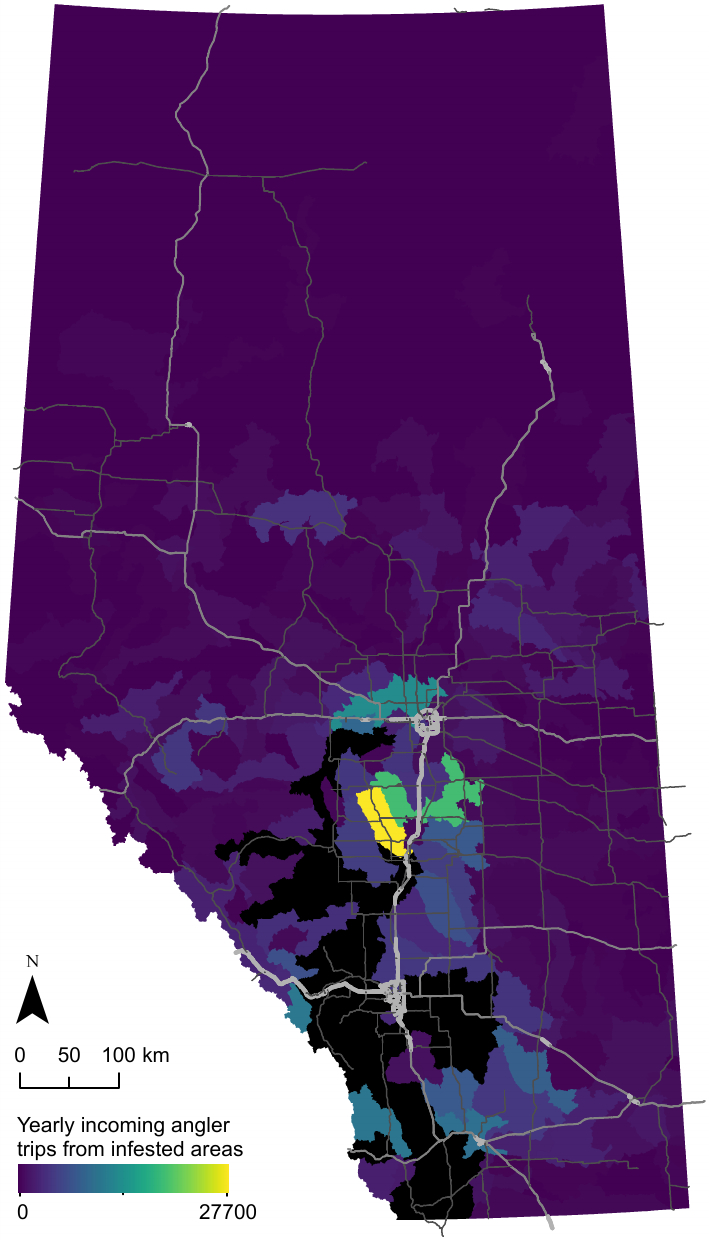}
			
		}\caption{(a) Number of consecutive trips to subbasin pairs and (b) total number
			of incoming trips by potentially infested anglers. In (a) only subbasin
			pairs with more than $100$ trips per year are shown. In (b) black
			colours depict subbasins that are already infested (March 2020).}
	\end{figure}

	Angler trips were estimated to be strongest between subbasins located
	close to metropolitan areas, with estimates up to $22.3$ thousand
	($95\%$ confidence interval $[14.5\e 3,32.6\e 3]$) directly consecutive
	angler trips per year (Figure~\ref{fig:HUC-HUC-trips}). The subbasin
	most at risk of receiving anglers infested with whirling disease propagules
	were those located close to larger cities and in proximity to the
	already infested area (Figure~\ref{fig:HUC-risk}). The subbasin
	with the highest inflow of high-risk trips was estimated to receive
	$27.7$ thousand ($[18.1\e 3,40.3\e 3]$) such trips per year. The
	estimated mean number of fishing days per angler and year was $20.5$
	($[12.5,34.1]$) as per our model. For comparison, the corresponding
	estimate from a $2016$ mail-out survey was $18$ (standard deviation
	between $0.9$ and $2.7$) \citep{dfo_survey_2019}. 
	
\begin{figure}
		\subfloat[]{\includegraphics[width=0.33\textwidth]{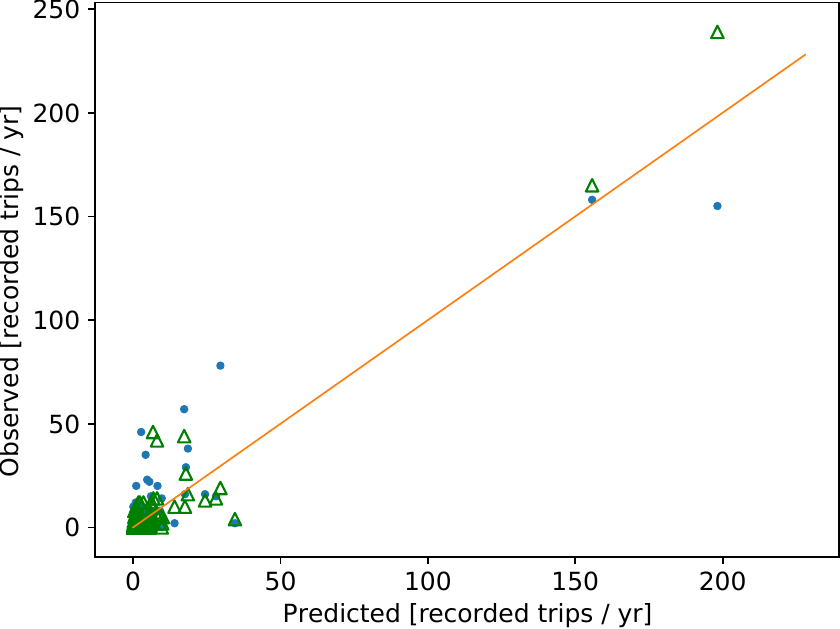}
			
		}\hspace*{\fill}\subfloat[]{\includegraphics[width=0.33\textwidth]{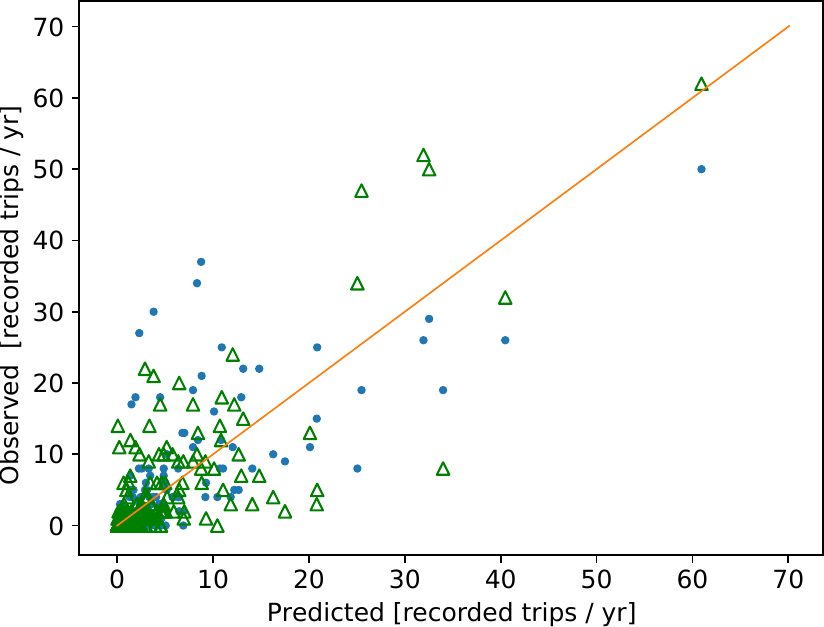}
			
		}\hspace*{\fill}\subfloat[]{\includegraphics[width=0.33\textwidth]{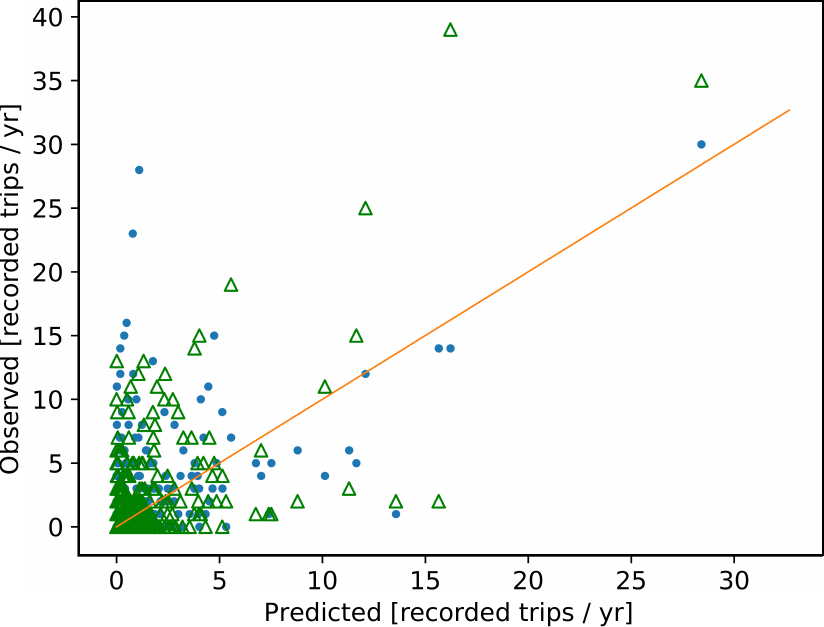}
			
		}
		
		\caption{Predicted and observed mean values of yearly recorded angler trips
			(a) by origin locality, (b) by destination subbasin, and (c) by locality-subbasin
			pair. The values used to fit the model are depicted as solid blue
			circles; the values computed from the independent fitting dataset
			are drawn as hollow green triangles. The orange line indicates where
			predictions and observations would coincide.\label{fig:Predicted-observed}}
		
	\end{figure}

	The pseudo-$R^{2}$ values of the model components decreased as more
	complexity was added. The submodel for the day suitability achieved
	a pseudo-$R^{2}$ of $0.47$. Adding the component for the origin-dependent
	trip rates and the destination choice probabilities yielded a value
	$0.35$. Predicted and observed values for this model component are
	depicted in Figure \ref{fig:Predicted-observed}. The joint model
	including the remaining choice parameters had a pseudo-$R^{2}$ of
	$0.26$. 
	
	The submodel for the traffic between origins and destinations estimated
	the outflow from origins with a mean error of $1.48$, the inflow
	to destinations with a mean error of $1.6$ and the traffic between
	individual pairs with a mean error of $0.0073$. The corresponding
	values obtained by using the data directly were $1.83$, $1.81$,
	and $0.0074$. That is, applying the model did not increase the error.
	
	\section{Discussion}
	
	Mobile apps exist for a variety of outdoor activities (e.g. birding,
	hiking and fishing) that could be related to the spread of animal
	diseases and invasive species. These apps can yield highly detailed
	individual-specific, spatially and temporally representative data
	and provide valuable insights into the traffic of anthropogenic vectors
	of invasive species and pathogens \citep{papenfuss_smartphones_2015,venturelli_angler_2017}.
	However, though the datasets collected via mobile apps can be large,
	they often cover only a small fraction of all trips of potential vectors,
	making direct estimates via the data's empirical distribution error-prone.
	Our results indicate that modelling approaches can reduce this issue
	and provide additional insights into the mechanisms behind vector
	movement. 
	
	Conversely, models can profit strongly from the new data source. 
	Though mail-out or online surveys could collect the same data as apps
	in principle, our modelling approach based on app data has the following
	advantages: 
	
	\paragraph*{(1) Increased accuracy. }
	
	Apps can be downloaded by users from different geographical areas,
	and data may be collected over extended time periods. Therefore, inference
	from app data is generally less sensitive to local and temporal peculiarities,
	and modellers can identify and account for the sources of spatial
	and temporal heterogeneity. This makes the estimates more accurate,
	especially when results are extrapolated into the future or to larger
	geographical scales. Furthermore, the temporal fingerprint of app
	data records permits a longitudinal study design.  By considering
	the day-to-day variations of the data, the unexplained recurring stochasticity
	in individual decisions can be distinguished from systemic errors
	due to misspecified models. Without this distinction (e.g. \citealp{drake_least-cost_2010,muirhead_prediction_2011,muirhead_evaluation_2011}),
	residuals would be solely attributed to the stochasticity in the individuals'
	decisions, and the dispersion parameter would be overestimated \citep{fischer_hybrid_2020}.
	Then, low-traffic angler flows gain an inordinate weight when the
	model is fitted to observations, resulting in decreased model accuracy
	(see Appendix~S6). 
	
	\paragraph*{(2) Estimates account for reduced vector mobility due to individual
		preferences. }
	
	Accounting for anglers' individual preferences  allowed us to estimate
	the frequency at which they switch destinations, potentially transporting
	propagules. We found that in $64\%$ of their trips, anglers revisit
	their previous destination and hence do not spread invasive species
	and pathogens to new areas. Furthermore, anglers tend to choose half
	of their fishing destinations from spatially contained areas. This
	suggests that models disregarding the correlations within anglers'
	destination choices (e.g. \citealp{bossenbroek_prediction_2001,leung_predicting_2004})
	are prone to overestimating traffic between distant destinations.

	\paragraph*{(3) Absolute estimates of vector traffic can be obtained without
		additional survey data.}
	
	It is difficult to obtain absolute traffic estimates from survey or
	app data without knowing which fraction of trips surveyed individuals
	or app users report. However, by considering anglers' tendency to
	revisit previous destinations, we were able to infer this missing
	information. If an angler does not record all their trips, the probability
	that the next trip they record has the same destination as their previously
	recorded trip decays with time, because they may make additional unrecorded
	trips to other destinations in the mean time. The slope at which the
	fraction of consecutively recorded trips with same destinations decays
	with the intermediate time depends on the trip recording probability.
	This makes it possible to infer this information from the data.
	
	Absolute estimates of traffic enable modellers to link traffic to
	invasion or infection success. This link is needed to predict the
	spread of a disease or invasive species \citep{lewis_mathematics_2016}.
	Though invasion success can be estimated based on relative traffic
	estimates if historical invasion data are available for the studied
	area \citep{leung_predicting_2004,muirhead_modelling_2006,potapov_models_2011},
	these estimates will remain site-specific unless the scaling of the
	traffic is known or the same traffic model is at the other site. Transferring
	a traffic model to a new site requires that similar data are available
	at the new site and that vectors behave the same. Absolute traffic
	estimates allow modellers to estimate the establishment success per
	individual vector and to transfer such information to or from other
	study areas.

	\subsection*{Validity of the estimates}
	
	Exploiting the connection between reported destinations and the completeness
	of the data, we estimated how many days an average angler goes fishing
	in Alberta per year. Our estimate for this quantity agreed with an
	independent estimate, differing only by about one standard deviation
	of the independent estimate. This suggests that our approach estimates
	the overall angler trip frequency accurately and hence can be used
	to obtain absolute traffic estimates even if the fraction of reported
	trips is unknown. Note, however, that the confidence interval for
	our estimate was relatively large and had an upper bound $60\%$ higher
	than the estimated value. Hence, additional surveys determining the
	total trip count -- and thereby the fraction of reported trips --
	remain worthwhile to reduce model uncertainty. We refrained from incorporating
	this additional information, because our estimate of the total trip
	counts was already in agreement with the independent estimate. 
	
	The predicted vs. observed analysis of the submodel for angler activeness
	and destination choice probabilities indicated a reasonable estimation
	accuracy. Comparison with the raw data showed that the model extrapolates
	the limited available data to all origin-destination pairs without
	introducing additional error -- the model did even slightly better
	than the direct estimates. This is due to the stochasticity inherent
	in the system and the limited number of available trip records. 
	
	The pseudo-$R^{2}$ values we computed for different model components
	decreased as more complexity was added. This is expected, because
	the level of detail of the validation data increased as additional
	model components were considered. The differing level of detail makes
	it difficult to compare our pseudo-$R^{2}$ values to similar metrics
	obtained in studies with less rich data sources (e.g. \citealp{drake_least-cost_2010,chivers_predicting_2012}).
	Nonetheless, the pseudo-$R^{2}$ value we reported may be a helpful
	benchmark for future studies on a similar system.

	\subsection*{Management implications}
	
	Our approach facilitates management in three ways: (1) it provides
	location-specific risk proxies; (2) it shows which locations are best
	connected and might thus be potential hubs for secondary infections;
	and (3) it helps to identify the origins of the agents most at risk
	spreading the disease. While risk estimates facilitate early detection
	and rapid response to new infections, the latter two points help targeting
	management actions to the subbasins and localities where they are
	most effective. 
	
	\paragraph*{(1) Risk proxies }
	
	The subbasins with the highest propagule inflow were those encompassing
	extended water areas with confirmed fish presence and located close
	by a population centre in proximity of an infected subbasin. Our results
	indicate that anglers show a strong preference for fishing destinations
	near their homes, which also agrees with earlier studies \citep{drake_least-cost_2010,papenfuss_smartphones_2015,fischer_hybrid_2020}.
	Both the inscribed radius of the regions of preference and the distance
	at which traffic decays by half were at about $30\,\text{km}$, suggesting
	this as main spatial scale of angler traffic. Besides water area and
	fish species confirmations, the number of campgrounds was another
	useful indicator of attractiveness, probably because they are typically
	built in scenic areas attractive to outdoor tourists.  
	
	The total waterbody circumference (i.e., shoreline) was positively
	connected with angler traffic as well, but the best-AIC model did
	not incorporate this covariate in conjunction with fish presence data,
	potentially due to incomplete species data in smaller rivers. Interestingly,
	the number of visits of waterbody-specific websites was not a worthwhile
	additional predictor for attractiveness -- perhaps because this number
	is also affected by the waterbodies' proximity to angler origins and
	their repulsiveness. 
	
	\paragraph*{(2) Potential hubs}
	
	Uninfected subbasins with strong connections to other uninfected subbasins
	might become significant hubs for secondary infections. This makes
	these well-connected subbasins a primary target for surveillance and
	rapid response measures. In our model, these subbasins have similar
	characteristics to those receiving most high-risk trips (see above)
	except for being located farther away from present infections.
	
	\paragraph*{(3) Most relevant angler origins}
	
	Education and outreach measures may be applied with different intensity
	in different localities. According to our results, maximal high-risk
	trip counts per inhabitant are found in rural areas, where the density
	of anglers per inhabitant is higher there than in cities. Hence, general
	outreach may be most effective in rural areas close to the edge of
	the infested area. In contrast, the number of high-risk trips per
	angler was also high in cities, making cities close to the infected
	area good soil for outreach measures specifically targeting anglers.
	Outreach may be cheaper and hence more cost-effective in cities than
	in rural areas. 
	
	\subsection*{Limitations and potential extensions}
	
	Voluntarily reported data such as app data can be prone to a variety
	of biases ranging from demographic bias to avidity bias \citep{venturelli_angler_2017}.
	For example, app usage may be higher among the young population from
	urbanized areas, and active sport fishers may use the app more frequently.
	As a result, certain demographic groups may be underrepresented, and
	highly active app users may influence the estimates disproportionately.
	This can lead to biased and over-confident results. 
	
	For the data from the MyCatch app used in this study, no spatial bias
	was detected \citep{johnston_comparative_2021}. This increases the
	credibility of our results. Nonetheless, some potential sources of
	bias remain, and other datasets may suffer from stronger bias. This
	issue could be addressed with additional data. If demographic and
	socioeconomic data are collected along with the other app data, these
	covariates could be directly incorporated into the model. If an independent
	sample with these data is available, it can be used to weigh the observations
	differently \citep{chen_doubly_2020}. 
	
	Our model uses socioeconomic covariates to estimate the activeness
	of anglers at different localities. Though incorporating population
	counts and the mean income in localities improved the model fit significantly,
	angler activeness is unlikely to be uniform within localities. Since
	there is no obvious mechanistic justification for the dependency of
	angler activeness on the covariates we used, we emphasize the phenomenological
	nature of the model and refrain from further analysis of the underlying
	mechanisms. 
	
	We also assume that the individual preferences of individuals do not
	change with time. This assumption may be inaccurate if extended time
	periods are considered, and individuals may change both their region
	of preference and their home place. As a result, the connectivity
	of close-by locations in directly consecutive trips might be underestimated.
	This issue could be addressed by splitting the dataset into subsets
	for different time periods and to treat them as independent replicates
	of one another.
	
	By fitting model components in separate steps, we obtained multiple
	estimates for the dispersion parameter $\alpha$. This indicates that
	some sources of stochasticity, such as weather, act on small scales
	only and have a reduced impact on an aggregate level. This is a common
	effect seen in ecological models \citep{dungan_balanced_2002} and
	difficult to resolve without ignoring spatial correlations completely
	or strongly increasing the model's complexity. However, since the
	estimates for the other parameters were insensitive to moderate variations
	of the dispersion parameter (Appendix~S7), our estimates for the mean angler
	traffic remain valid despite the scale-dependency of the dispersion
	parameter. 
	
	Noting that propagules may be washed out at any site visited by anglers,
	we concentrated on estimating the number of consecutive fishing trips
	to distinct subbasins. This is in line with existing studies on invasive
	species transport \citep{bossenbroek_prediction_2001,leung_predicting_2004,potapov_models_2011,muirhead_evaluation_2011}.
	An alternative approach is to consider all trips that anglers make
	within the time frame propagules may survive \citep{papenfuss_smartphones_2015}.
	Note, however, that we assumed that anglers choose their destinations
	independently of the past unless they revisit their previous destination.
	Therefore, the number of higher-order trips between two subbasins
	equals the count of directly consecutive trips unless a constrained
	time frame is considered. Thus, incorporating higher-order trips comes
	down to determining a sensible time scale for propagule decay. 
	
	We focused on intra-provincial angler traffic, but our model could
	be extended to also consider non-resident anglers visiting the province
	on vacation trips. Within our model framework, such anglers could
	be added to the populations of the localities where they reside temporally.
	However, in Alberta, non-resident anglers are responsible for only
	$2.6\%$ of the angler traffic (measured in yearly fishing days) \citep{dfo_survey_2019},
	which is below the error margin of our estimates.
	
	\subsection*{Alternative approaches}
	
	Our model's complexity is driven by the challenges stemming from missing
	trip records in app data. If complete trip records were available,
	a phenomenological gravity model for the connectedness of destinations
	\citep[e.g.][]{potapov_stochastic_2010,muirhead_evaluation_2011,chivers_predicting_2012}
	could be constructed directly. However, complete trip records are
	often difficult to obtain without strong simplifying assumptions,
	such as that vectors have visited all distinations they reported with
	the same frequency \citep{potapov_stochastic_2010}. 
	
	If data on the temporal progression of the disease or invasion are
	available, the connectedness between destinations could also be inferred
	based on the observed infestation dynamics \citep{bossenbroek_prediction_2001,leung_predicting_2004}.
	However, the resulting traffic estimates can typically not be validated
	due to the lack of data. Besides, this approach makes it necessary
	that an establishment model exists that accounts for on-site conditions
	affecting establishment success. 
	
	\section{Conclusion}
	
	The increasingly widespread use of mobile apps by anglers, hikers,
	campers, and other potential vectors of invasive species and pathogens
	opens new opportunities for research and management. To exploit the
	full potential of this new data source, models accounting for spatial,
	temporal, and individual heterogeneity are needed. The presented model
	demonstrates the wealth of information that can be gained from app
	data, including (1) temporally explicit estimates of vector traffic
	between cities and waterbodies, (2) estimates of how often vectors
	choose new trip destinations, potentially carrying propagules to other
	places, and (3) the spatial scale at which individual local preferences
	play a major role in vectors' decisions. Our results suggest that
	ignoring individual-specific components in vectors' decision making
	can bias estimates by underestimating local traffic and overestimating
	long-distance traffic. We furthermore showed that incorporating vectors'
	tendency to revisit past locations can bridge the common data gap
	arising from incomplete trip records reported by app users.

	\section*{Acknowledgements}
	
	All authors are thankful for the funding received from Alberta Environment
	and Parks. In addition, PR and MAL thankfully acknowledge an NSERC
	Discovery Grant and MAL also gratefully acknowledges a Canada Research
	Chair. All authors would like to thank the anglers who voluntarily
	provided data via the MyCatch app and the Anglers' Atlas website.
	Furthermore, the authors thank the Goldstream Publishing staff, who
	helped prepare the data for the analysis as well as the members of
	the Lewis Research Group, who contributed through helpful feedback
	and discussions.
	
	\section*{Conflict of interest}
	
	Sean Simmons is founder and president at Angler's Atlas and MyCatch. The authors declare no conflict of interest.

	\section*{Author contributions}
	
	Pouria Ramazi, Sean Simmons, Mark S. Poesch, and Mark A.
	Lewis conceived the project. Sean Simmons collected and prepared
	the app data and the website data. All authors contributed to the methods;
	Samuel M. Fischer finalized and implemented the model. Samuel M. Fischer 
	and Pouria Ramazi led the writing
	of the manuscript. All authors contributed critically to the drafts
	and gave final approval for publication.

	\section*{Data Availability}
	
	The sources of the data used in this study are listed below. The
	code used in this study can be found at 
	\href{https://github.com/vemomoto/indspecvemo}{github.com/vemomoto/indspecvemo}.
	The compiled dataset used in this study is available via the Dryad Digital Repository \href{https://doi.org/10.5061/dryad.6m905qg3j}{doi.org/10.5061/dryad.6m905qg3j}
	\citep{fischer_data_2022}.

	\section*{Data Sources\label{sec:Data-sources}}
	
	\footnotesize \renewcommand{\arraystretch}{1.2} \setlength\arrayrulewidth{0.5pt}%
	\begin{tabular}{>{\raggedright}p{0.28\textwidth}>{\raggedright}p{0.22\textwidth}>{\raggedright}p{0.42\textwidth}}
		\toprule 
		\textbf{Data} & \textbf{Source} & \textbf{URL}\tabularnewline
		\midrule
		\midrule 
		Angler App Data & \multirow{4}{0.22\textwidth}{Goldstream Publishing} & \multirow{4}{0.42\textwidth}{The raw data are not available online. A compilation of the data used
			as input for the model will be made available along with the article.}\tabularnewline
		\cmidrule{1-1} 
		Website Visit Data &  & \tabularnewline
		\cmidrule{1-1} 
		Species Vote Data &  & \tabularnewline
		\cmidrule{1-1} 
		Waterbody GIS Data &  & \tabularnewline
		\midrule 
		Subbasin GIS Data & Government of Alberta (Alberta Environment and Parks) &
		Original URL (not available anymore):
		 \href{https://maps.alberta.ca/genesis/rest/services/Hydrologic_Unit_Code_Watersheds_of_Alberta/}{maps.alberta.ca/genesis/rest/services/\\Hydrologic\_Unit\_Code\_Watersheds\_of\_Alberta} Potential alternative data source: \href{https://geospatial.alberta.ca}{geospatial.alberta.ca}
		 \tabularnewline
		\midrule 
		Locality GIS Data & Open Street Map & \href{https://www.openstreetmap.org/}{www.openstreetmap.org}\tabularnewline
		\midrule 
		Campground GIS Data & USCAmpgrounds & \href{http://www.uscampgrounds.info/}{www.uscampgrounds.info}\tabularnewline
		\midrule 
		Angler License Count Data & Government of Alberta (Alberta Environment and Parks) & Retrieved through a standard data request from the Fish and Wildlife
		Management Information System \href{\%20https://www.alberta.ca/access-fwmis-data.aspx}{www.alberta.ca/access-fwmis-data.aspx}\tabularnewline
		\midrule 
		Population Count Data  & Government of Alberta & \href{https://open.alberta.ca/opendata/alberta-municipal-affairs-population-list}{open.alberta.ca/opendata/alberta-municipal-affairs-population-list}\tabularnewline
		\midrule 
		Population Income Data & Government of Alberta & \href{https://open.alberta.ca/dataset/labour-income-profile-for-all-forward-station-areas-city-totals-and-rural-postal-codes-canada-2013}{open.alberta.ca/dataset/labour-income-profile-for-all-forward-station-areas-city-totals-and-rural-postal-codes-canada-2013}\tabularnewline
		\bottomrule
	\end{tabular}

	\bibliographystyle{mee}
	
\end{document}


\title{\vspace{-1cm}Supporting Information for ``Boosting Propagule Transport Models with Individual-Specific Data
from Mobile Apps''}

\maketitle

\renewcommand{\theequation}{S\arabic{equation}}
\renewcommand{\thefigure}{S\arabic{figure}}
\renewcommand{\thetable}{S\arabic{table}}
\renewcommand{\thesection}{S\arabic{section}}


\vspace{-1.5cm}

\section{Overview of mathematical symbols used in the paper\label{sec:notation-table}}

The table below provides an overview of the mathematical symbols used
in the main text.

{\footnotesize \setlength\arrayrulewidth{0.5pt}%
	\begin{longtable}[c]{>{\raggedright}p{1.4cm}>{\raggedright}p{15cm}}
		\toprule 
		\textbf{Symbol} & \textbf{Explanation}\tabularnewline
		\midrule
		\hline
		\endfirsthead
		\toprule 
		\textbf{Symbol} & \textbf{Explanation}\tabularnewline
		\midrule
		\hline
		\endhead
		$i$ & Origin locality\tabularnewline
		$j$ & Destination subbasin\tabularnewline
		$t$ & Day\tabularnewline
		$\mathcal{R}$  & Region of preference; a set of subbasins\tabularnewline
		\midrule 
		$N_{t}$  & Total number of recorded angler trips on day $t$ \tabularnewline
		$N_{ijt}$  & Total number of recorded angler trips from $i$ to $j$ on day $t$ \tabularnewline
		$N_{ij,\mt{fit}}$ & Total number of recorded angler trips from $i$ to $j$ in the training
		dataset\tabularnewline
		$N_{ij,\mt{val}}$ & Total number of recorded angler trips from $i$ to $j$ in the validation
		dataset\tabularnewline
		$\bar{N}_{ij,\mt{fit}}$ & Empirical estimate for the \emph{annual} number of recorded angler
		trips from $i$ to $j$ based on the training data\tabularnewline
		$\bar{N}_{ij,\mt{val}}$ & Empirical estimate for the \emph{annual} number of recorded angler
		trips from $i$ to $j$ based on the validation data\tabularnewline
		\midrule 
		$n_{i}$  & Number of anglers residing at locality $i$\tabularnewline
		$\tilde{n}_{i}$  & Number of app-using anglers residing at locality $i$\tabularnewline
		\midrule 
		$T$  & Number of days in the study period\tabularnewline
		$\mathfrak{R}$ & Set of all potential regions of preference\tabularnewline
		$\mathcal{X}$ & Group of origin or destination covariates that jointly lead to high
		traffic; a set of covariates\tabularnewline
		$\bm{x}$  & Covariate; vector of covariate values\tabularnewline
		\midrule 
		$\mu_{i}$  & Mean \emph{daily} number of trips \emph{per angler} from origin $i$\tabularnewline
		$\bar{\mu}$ & Mean total\emph{ daily} number of recorded trips\tabularnewline
		$\mu_{j_{1}j_{2}}$ & Mean total\emph{ yearly} number of trips anglers make to destination
		$j_{1}$ directly after visiting destination $j_{2}$\tabularnewline
		$\mu_{ij_{1}j_{2}}$ & Mean total\emph{ yearly} number of directly consecutive trips from
		origin $i$ to destinations $j_{1}$ and $j_{2}$ \tabularnewline
		\midrule 
		$p_{ij}$  & Probability that an angler from origin $i$ visits destination $j$\tabularnewline
		$p_{i\mathcal{R}}$  & Probability that an angler from origin $i$ has $\mathcal{R}$ as
		region of preference \tabularnewline
		$p_{ij|\mathcal{R}}$  & Probability that an angler from origin $i$ with region of preference
		$\mathcal{R}$ visits destination $j$\tabularnewline
		\midrule 
		$\varepsilon_{t}$\emph{ } & Suitability of day $t$ for going fishing\tabularnewline
		$\tau_{t}$ & Expected suitability of day $t$\tabularnewline
		\midrule 
		$\alpha$ & Dispersion parameter\tabularnewline
		$\psame$ & Probability to revisit the previous destination\tabularnewline
		$\pregion$ & Probability to constrain the destination choice to the region of preference\tabularnewline
		$\papp$ & Probability to use the app\tabularnewline
		$\preport$ & Probability to record a trip\tabularnewline
		\midrule 
		$a_{j}$ & Attractiveness of destination $j$\tabularnewline
		$D$  & Distance decay function for the destination choice probabilities \tabularnewline
		$f_{\mt{vM}}$ & Probability density function of the von Mises distribution\tabularnewline
		\midrule 
		$c$  & Scaling constant for the mean number of trips per day\tabularnewline
		$C$ & Scaling constant for the mean number of \emph{recorded} trips per
		day\tabularnewline
		$c_{\mt{norm}}$  & Normalization constant for the mean day suitability\tabularnewline
		\midrule 
		$\rho$  & Radius of the inscribed circle of regions of preference\tabularnewline
		$\delta_{0}$ & Distance of half choice-probability decay\tabularnewline
		$c_{\mt{week}}$ & Week addition constant\tabularnewline
		$\theta_{\mt{week}}$ & Week location constant\tabularnewline
		$\kappa_{\mt{week}}$ & Week shape constant\tabularnewline
		$c_{\mt{year}}$ & Year addition constant\tabularnewline
		$\theta_{\mt{year}}$ & Year location constant\tabularnewline
		$\kappa_{\mt{year}}$ & Year shape constant\tabularnewline
		\midrule 
		$\beta_{\bm{x}}$  & Scaling parameter for covariate $\bm{x}$ \tabularnewline
		$\gamma_{\bm{x}}$ & Power parameter for covariate $\bm{x}$ \tabularnewline
		\midrule 
		$\omega_{\mt{fit}}$  & Fraction of the data used for fitting the model\tabularnewline
		$\omega_{\mt{val}}$  & Fraction of the data used for model validation\tabularnewline
		\midrule 
		$\delta_{j_{1}j_{2}}$  & Kronecker delta; $1$ if $j_{1}=j_{2}$ and $0$ otherwise\tabularnewline
		\bottomrule
\end{longtable}}

\section{Computing the expected number of trips between angler destinations\label{sec:APX-mean-HUC-HUC}}

Below, we show how the mean number of consecutive trips to two destinations
can be computed efficiently.  As described in the main text, the
mean number of consecutive trips to two destinations $j_{1}$ and
$j_{2}$ by anglers from origin $i$ is given by
\begin{multline}
	\mu_{ij_{1}j_{2}}=n_{i}\mu_{i}T\Bigg(\psame\delta_{j_{1}j_{2}}p_{ij_{1}}+\left(1-\psame\right)\Bigg(\pall^{2}p_{ij_{1}}p_{ij_{2}}+\pall\pregion p_{ij_{1}}\smo{\mathcal{R}\,:\,j_{2}\in\mathcal{R}}p_{i\mathcal{R}}p_{ij_{2}|\mathcal{R}}\\
	+\pregion\pall p_{ij_{2}}\smo{\mathcal{R}\,:\,j_{1}\in\mathcal{R}}p_{i\mathcal{R}}p_{ij_{1}|\mathcal{R}}+\pregion^{2}\smo{\mathcal{R}\,:\,j_{1},j_{2}\in\mathcal{R}}p_{i\mathcal{R}}p_{ij_{1}|\mathcal{R}}p_{ij_{2}|\mathcal{R}}\Bigg)\Bigg).\label{eq:mean_HUC_HUC_traffic-APX}
\end{multline}
After expanding and simplifying equation (\ref{eq:mean_HUC_HUC_traffic-APX}),
we obtain 
\begin{alignat}{1}
	\mu_{ij_{1}j_{2}}= & n_{i}\mu_{i}T\Bigg(\psame\delta_{j_{1}j_{2}}p_{ij_{1}}+\left(1-\psame\right)\Bigg(\pall^{2}p_{ij_{1}}p_{ij_{2}}+\pall\pregion\smo{\mathcal{R}\,:\,j_{2}\in\mathcal{R}}\frac{p_{ij_{1}}p_{ij_{2}}}{\smo{\tilde{\mathcal{R}}}\smo{j\in\tilde{\mathcal{R}}}p_{ij}}\nonumber \\
	& +\pall\pregion p_{ij_{2}}\smo{\mathcal{R}\,:\,j_{1}\in\mathcal{R}}\frac{p_{ij_{1}}p_{ij_{2}}}{\smo{\tilde{\mathcal{R}}}\smo{j\in\tilde{\mathcal{R}}}p_{ij}}+\pregion^{2}\smo{\mathcal{R}\,:\,j_{1},j_{2}\in\mathcal{R}}\frac{p_{ij_{1}}p_{ij_{2}}}{\smo{j\in\mathcal{R}}p_{ij}\cdot\smo{\tilde{\mathcal{R}}}\smo{j\in\tilde{\mathcal{R}}}p_{ij}}\Bigg)\Bigg)\nonumber \\
	= & n_{i}\mu_{i}Tp_{ij_{1}}\Bigg(\psame\delta_{j_{1}j_{2}}+\left(1-\psame\right)p_{ij_{2}}\Bigg(\pall^{2}+\frac{1}{\smo{\tilde{\mathcal{R}}}\smo{j\in\tilde{\mathcal{R}}}p_{ij}}\Bigg(\nonumber \\
	& \qquad\pall\pregion\left(\left|\left\{ \mathcal{R}\,:\,j_{1}\in\mathcal{R}\right\} \right|+\left|\left\{ \mathcal{R}\,:\,j_{2}\in\mathcal{R}\right\} \right|\right)+\pregion^{2}\smo{\mathcal{R}\,:\,j_{1},j_{2}\in\mathcal{R}}\frac{1}{\smo{j\in\mathcal{R}}p_{ij}}\Bigg)\Bigg)\Bigg),\label{eq:mean_HUC_HUC_traffic-simplified}
\end{alignat}
where $\left|\cdot\right|$ denotes the counting norm. The right hand
side of (\ref{eq:mean_HUC_HUC_traffic-simplified}) can be evaluated
efficiently for all origins $i$ and destinations $j_{1}$ and $j_{2}$
if the individual terms are stored for reuse. 

\section{Likelihood derivation\label{sec:APX-Likelihood}}

Here, we show how the likelihood functions used in the model fitting
stages are derived and how they can be computed efficiently.

\subsection{Day suitability\label{subsec:APX-Likelihood-Time}}

In this section, we derive the likelihood function for the distribution
of trips recorded each day $N_{t}$. Recall that the rate of trips
for anglers from origin $i$ on day $t$ is assumed to be $\mu_{i}\varepsilon_{t}$.
As a result, the number of trips that such anglers make on day $t$
is Poisson distributed with mean $\mu_{i}\varepsilon_{t}$. Since
$\varepsilon_{t}$ follows a gamma distribution with shape parameter
$\frac{\tau_{t}}{\alpha}$ and scale parameter $\alpha$, the distribution
of the total number $\bar{N}_{t}$ of trips on day $t$ is 
\begin{align}
	\bar{N}_{t} & \sim\smo i\smft{k=1}{n_{i}}\poissondist{\mu_{i}\varepsilon_{t}}\nonumber \\
	& =\smo i\poissondist{n_{i}\mu_{i}\varepsilon_{t}}\nonumber \\
	& =\poissondist{\varepsilon_{t}\smo in_{i}\mu_{i}},
\end{align}
which in turn is a negative binomial distribution with dispersion
parameter $\frac{\alpha}{\tau_{t}}$ and mean $\tau_{t}\smo in_{i}\mu_{i}$
\citep{villa_using_2006}. 

When computing the likelihood of our observed data, we need to take
into account that many anglers are not using the app. Consequently,
we need to use the number $\tilde{n}_{i}$ of app users in locality
$i$ instead of the angler number $n_{i}$. As $\tilde{n}_{i}$ is
a random variable itself, we were not able to derive a closed-form
solution of the exact likelihood function. However, since we consider
the \emph{total }number of trips each day, the identities of the individual
anglers who made the trips is of minor importance, and we may consider
the individual trips on a day as independent from one another. This
approximation is appropriate, because we consider a large number of
anglers and a relatively small number of trips. The number of trips
recorded via the app is then 
\begin{equation}
	N_{t}\sim\binomdist{\bar{N}_{t}}{\papp\preport}.
\end{equation}
This can be expressed as a negative binomial distribution with dispersion
parameter $\frac{\alpha}{\tau_{t}}$ and mean $\tau_{t}\bar{\mu}$,
where $\bar{\mu}=\papp\preport\smo in_{i}\mu_{i}$ \citep{villa_using_2006}.

Let $k_{t}$ denote the number of trips recorded on day $t$. Following
the derivation above, the likelihood function then reads
\begin{equation}
	L_{\mt{time}}=\pro t\binom{k_{t}+\frac{\tau_{t}}{\alpha}-1}{\frac{\tau_{t}}{\alpha}-1}\left(\frac{1}{1+\alpha\bar{\mu}}\right)^{\frac{\tau_{t}}{\alpha}}\ap{\frac{\alpha\bar{\mu}}{1+\alpha\bar{\mu}}}^{k_{t}}.
\end{equation}
The log-likelihood function (disregarding an irrelevant constant shifting
term) is 
\begin{equation}
	\ell_{\mt{time}}=\smo t\ln\Gamma\ap{k_{t}+\frac{\tau_{t}}{\alpha}}-\smo t\ln\Gamma\ap{\frac{\tau_{t}}{\alpha}}+K\left(\ln\alpha+\ln\bar{\mu}\right)-\left(\frac{T}{\alpha}+K\right)\ln\ap{1+\alpha\bar{\mu}},\label{eq:l-time}
\end{equation}
where $T=\smo t\tau_{t}$ is the number of days in the study period,
and $K=\smo tk_{t}$ is the total number of observed trips. Note that
days $t$ with $k_{t}=0$ can be omitted in the sums in (\ref{eq:l-time})
without changing $\ell_{\mt{time}}$. 

\subsection{Angler activeness and destination choice probabilities\label{subsec:APX-Likelihood-Gravity}}

The likelihood function for the stochastic gravity model yielding
the angler activeness and the destination choice probabilities can
be derived analogously to the likelihood function for the day suitability.
Again, we consider individual trips on a day as independent from one
another to improve the computational performance. Then, the number
of trips between each origin-destination pair is negative binomially
distributed. Since it is difficult to account for the stochastic dependency
of angler counts on the same day, we also assume that the traveller
counts between each origin and destination are mutually independent. 

To derive the likelihood function, let $\check{\mu}_{i}=\mu_{i}/c$,
where $\mu_{i}$ and $c$ are defined as in equation (9)
(main text). The mean number of recorded trips from origin $i$ to
destination $j$ on day $t$ is 
\begin{align}
	\mu_{ijt} & =\text{\ensuremath{\tau_{t}}}n_{i}\mu_{i}p_{ij}\papp\preport\nonumber \\
	& =C\text{\ensuremath{\tau_{t}}}n_{i}\check{\mu}_{i}p_{ij}.
\end{align}
Here, $\text{\ensuremath{\tau_{t}}}$ and $n_{i}$ are already estimated
or known, $\tilde{\mu}_{i}$ and $p_{ij}$ are functions of parameters
and covariates (see equations (7)-(8)
in the main text), and $C=\papp\preport c$ is the product of not
individually identifiable parameters. The dispersion parameter of
the distribution is $\frac{\alpha}{\tau_{t}}$. With $k_{ijt}$ being
the number of recorded trips from origin $i$ to destination $j$
on day $t$, we obtain the likelihood function 
\begin{equation}
	L_{\mt{gravity}}=\pro{i,j,t}\binom{k_{ijt}+\frac{\tau_{t}}{\alpha}-1}{\frac{\tau_{t}}{\alpha}-1}\left(\frac{1}{1+\alpha Cn_{i}\check{\mu}_{i}p_{ij}}\right)^{\frac{\tau_{t}}{\alpha}}\ap{\frac{\alpha Cn_{i}\check{\mu}_{i}p_{ij}}{1+\alpha Cn_{i}\check{\mu}_{i}p_{ij}}}^{k_{ijt}}
\end{equation}
and the log-likelihood function (up to a constant shifting term)
\begin{align}
	\ell_{\mt{gravity}}= & \smo{i,j,t}\ln\Gamma\ap{k_{ijt}+\frac{\tau_{t}}{\alpha}}-\smo{i,j,t}\ln\Gamma\ap{\frac{\tau_{t}}{\alpha}}+\smo{i,j,t}k_{ijt}\ln\left(\alpha Cn_{i}\check{\mu}_{i}p_{ij}\right)\nonumber \\
	& \qquad-\smo{i,j,t}\left(\frac{\tau_{t}}{\alpha}+k_{ijt}\right)\ln\ap{1+\alpha Cn_{i}\check{\mu}_{i}p_{ij}}\\
	= & \smo{i,j,t\,:\,k_{ijt}\neq0}\ln\Gamma\ap{k_{ijt}+\frac{\tau_{t}}{\alpha}}-\smo tk_{t}\ln\Gamma\ap{\frac{\tau_{t}}{\alpha}}-\frac{T}{\alpha}\smo{i,j}\ln\ap{1+\alpha Cn_{i}\check{\mu}_{i}p_{ij}}\nonumber \\
	& +\smo{i,j\,:\,k_{ij}\neq0}k_{ij}\ln\left(\alpha Cn_{i}\check{\mu}_{i}p_{ij}\right)-\smo{i,j,t\,:\,k_{ijt}\neq0}k_{ijt}\ln\ap{1+\alpha Cn_{i}\check{\mu}_{i}p_{ij}}.
\end{align}
Here, $k_{ij}=\smo tk_{ijt}$ is the total number of trips observed
from origin $i$ to destination $j$. The time required to evaluate
$\ell_{\mt{gravity}}$ is dominated by either the number of origin-destination
pairs or the number of recorded trips.

\subsection{Choice parameters\label{subsec:APX-Likelihood-Choice-Parameters}}

To derive the likelihood for the remaining choice parameters, we express
the negative binomial probability mass function in a simpler way in
terms of the parameters $q=\frac{1}{1+\bar{\alpha}\mu}$, describing
the distribution's mean-to-variance ratio, and the parameter $r=\frac{1}{\bar{\alpha}}$,
which is proportional to the distribution's expected value. Here,
$\mu$ refers to the distribution's expected value and $\bar{\alpha}$
to respective dispersion parameter. With the changed parameterization,
the probability mass function of the negative binomial distribution
reads
\begin{equation}
	f(k;r,q)=\binom{k+r-1}{r-1}q^{r}\ap{1-q}^{k}.
\end{equation}

In contrast to the previous section, we need to consider the decisions
of each angler individually, to fit the remaining parameters. Consider
an angler from origin $i$ who uses the app and recorded trips on
days $t_{1},\dots,t_{m}$ to the destinations $j_{1},\dots,j_{m}$.
To compute the likelihood for this sequence of trips, we first determine
the probability that the angler records trips on the days $t_{1},\dots,t_{m}$.
Then we compute the probability that they choose the recorded locations
$j_{1},\dots,j_{m}$ given that they recorded trips on days $t_{1},\dots,t_{m}$. 

Let $k_{t}$ be the number of trips that the angler has made on day
$t$, and define $\tilde{q}_{i}=1/\left(1+\alpha\preport\mu_{i}\right)$
and $r_{t}=\frac{\tau_{t}}{\alpha}$. The probability that we observe
the given sequence of trips by the angler (ignoring the destinations)
is 
\begin{equation}
	\pro t\binom{k_{t}+r_{t}-1}{r_{t}-1}\tilde{q}_{i}^{r_{t}}\ap{1-\tilde{q}_{i}}^{k_{t}},\label{eq:time-sequence-pmf}
\end{equation}
which can be written as 
\begin{equation}
	\tilde{q}_{i}^{T/\alpha}\pro{t\,:\,k_{t}\neq0}\binom{k_{t}+r_{t}-1}{r_{t}-1}\ap{1-\tilde{q}_{i}}^{k_{t}}.
\end{equation}

Now we proceed with the second step and derive the probability for
a particular sequence of destination choices given the dates of the
corresponding trips. We have to take into account that the angler
may have made additional intermediate trips between any consecutive
pair of recorded trips. We start by deriving the probability $p_{ij_{l}j_{l+1}|\mathcal{R},k}$
that an angler from locality $i$ with preferred fishing region $\mathcal{R}$
records a trip to site $j_{l+1}$, given that (1) the trip they recorded
had the destination $j_{l}$ and (2) that they took $k-1$ trips between
the recorded trips.  Recall that on each trip, the angler can either
(A) refuse to choose a new destination and choose the last destination
again, (B) restrict the destination choice to their region of preference
$\mathcal{R}$, or (C) make an unrestricted choice from all available
destinations. Now note the following:
\begin{enumerate}
	\item The probability that the angler refuses to choose a new destination
	for all $k$ trips is $\delta_{j_{l}j_{l+1}}\psame^{k}$. Recall that
	$\delta_{j_{l}j_{l+1}}$ is $1$ if $j_{l}=j_{l+1}$ and $0$ else.
	The delta multiplier is necessary, because the probability that the
	angler chooses the same destination for all $k$ intermediate trips
	is $0$ if the first and the final destination are not equal.
	\item The probability that the angler chooses a new destination for at least
	one of the $k$ trips is $1-\left(1-\delta_{1l}\right)\psame^{k}$.
	Here, the term $1-\delta_{1l}$ models that on their very first trip,
	the angler cannot choose the same destination as on a previous trip. 
	\item Suppose the angler chooses a new destination for at least one of the
	$k$ trips and consider among these trips the last one for which the
	angler chooses a new destination. The probability that they make an
	unconstrained destination choice and choose $j_{l+1}$ as destination
	on this trip is $\pall p_{ij_{l+1}}$.\label{enu:point-all}
	\item Under the same conditions as described in point \ref{enu:point-all},
	the probability that -- on the last trip where they choose a new
	destination -- the angler restricts their destination choice to their
	preferred region and chooses $j_{l}$ is $\I{\mathcal{R}}{j_{l+1}}\pregion p_{ij_{l+1}|\mathcal{R}}$.
	Here, $\I{\mathcal{S}}s$ is the indicator function, which is $1$
	if $s\in\mathcal{S}$ and $0$ else.
\end{enumerate}
Putting the above observations together, we obtain

\begin{equation}
	p_{ij_{l}j_{l+1}|\mathcal{R},k}=\left(1-\left(1-\delta_{1l}\right)\psame^{k}\right)\left(\pall p_{ij_{l+1}}+\I{\mathcal{R}}{j_{l+1}}\pregion p_{ij_{l+1}|\mathcal{R}}\right)+\delta_{j_{l}j_{l+1}}\psame^{k}.
\end{equation}

Since $p_{ij_{l}j_{l+1}|\mathcal{R},k}$ depends on the (unknown)
number of intermediate trips between two recorded trips, we need to
derive the distribution of this number. Let $\mathcal{T}=\left\{ t\in\N\,:\,t_{l}<t<t_{l+1}\right\} $
be the set of days between the recorded trips $l$ and $l+1$. If
$l=0$, set $t_{0}$ as the last day before the study period started.
The number of trips that the angler makes during the days in $\mathcal{T}$
is negative binomially distributed with the parameters 
\begin{align}
	\bar{r}_{\mathcal{T}} & =\frac{1}{\alpha}\smo{t\in\mathcal{T}}\tau_{t}\\
	q_{i} & =\frac{1}{1+\alpha\mu_{i}}.
\end{align}
The distribution of unrecorded trips within the days in $\mathcal{T}$
is described by 
\begin{multline}
	\pr{k\text{ trips during }\mathcal{T}\,|\,\text{no trip recorded}}=\frac{\pr{k\text{ trips during }\mathcal{T}}\left(1-\preport\right)^{k}}{\pr{\text{no trips recorded during }\mathcal{T}}}\\
	=\binom{k+\bar{r}_{\mathcal{T}}-1}{\bar{r}_{\mathcal{T}}-1}\left(1-\left(1-\preport\right)\left(1-q_{i}\right)\right)^{\bar{r}_{\mathcal{T}}}\left(\left(1-\preport\right)\ap{1-q_{i}}\right)^{k}.\label{eq:dist-unreported-between}
\end{multline}
The factor $\left(1-\left(1-\preport\right)\left(1-q_{i}\right)\right)^{\bar{r}_{\mathcal{T}}}$
normalizes the distribution so that the probabilities for all $k\in\N$
sum to $1$. 

The distribution given by (\ref{eq:dist-unreported-between}) describes
the number of unrecorded trips during \emph{complete} days between
two trip records. However, multiple trips can be made on the same
day. In particular, the considered angler could have made unrecorded
intermediate trips on the days $t_{l}$ and $t_{l+1}$. Though multiple
trips on the same day may occur rarely only, overdispersed distributions
have the property that the probability for a second trip (given that
a first trip has been made) is higher than the unconditional probability
for a first trip. Therefore, ignoring the possibility of unrecorded
trips on the days $t_{l}$ and $t_{l+1}$ would skew the likelihood
significantly. 

Below, we will first derive the distribution of unrecorded trips on
a day where a trip has already been recorded. Afterwards, we determine
the joint distribution of unrecorded trips, before we finally incorporate
the probabilities to choose specific locations and determine the likelihood
function. Consider a day $t$ on which the angler makes $M_{t}\geq N_{t}$
trips, where they record $N_{t}\geq1$ of these trips. Partition the
day into time intervals separated by the recorded trips and consider
any of the arising sections. Let $k$ be the number of unrecorded
trips made during this day section. The number of possibilities to
distribute the remaining unrecorded trips over the other, remaining
sections is 
\begin{equation}
	\binom{M_{t}-k-1}{N_{t}-1},
\end{equation}
as can be computed with the stars and bars method. As a result, the
probability that the angler makes $k$ unrecorded trips in this time
interval is 
\begin{equation}
	\pr{k\text{ unrecorded trips in section}\,|\,\text{\ensuremath{M_{t}} total trips; \ensuremath{N_{t}} recorded trips}}=\frac{\binom{M_{t}-k-1}{N_{t}-1}}{\binom{M_{t}}{N_{t}}}.
\end{equation}
The normalization factor was obtained by summing over all $k\in\N$.

The number of trips recorded on the considered day follows a binomial
distribution with parameters $M$ and $\preport$; the total number
$M$ of trips follows a negative binomial distribution with parameters
$r_{t}=\frac{\tau_{t}}{\alpha}$ and $q_{i}=\frac{1}{1+\alpha\mu_{i}}$.
Hence, the joint probability that an angler makes $k$ unrecorded
trips between two recorded trips if they recorded $N_{t}$ trips in
total is 
\begin{align}
	& \pr{k\text{ unrecorded trips on day }t\,|\,\text{\ensuremath{N_{t}} trips recorded on day }t}\nonumber \\
	= & \bar{C}\smft{M_{t}=N_{t}+k}{\infty}\descr{\binom{M_{t}+r_{t}-1}{r_{t}-1}q_{i}^{r_{t}}\left(1-q_{i}\right)^{M_{t}}}{\pr{\text{make \ensuremath{M_{t}} trips}}}\descr{\binom{M_{t}}{N_{t}}\preport^{N_{t}}\left(1-\preport\right)^{m-N_{t}}}{\pr{\text{record \ensuremath{N_{t}} out of \ensuremath{M_{t}} trips}}}\descr{\binom{M_{t}-k-1}{N_{t}-1}\bigg/\binom{M_{t}}{N_{t}}}{\pr{\text{make \ensuremath{k} unreported trips}}}\nonumber \\
	= & \tilde{C}\smft{M_{t}=N_{t}+k}{\infty}\binom{M_{t}+r_{t}-1}{r_{t}-1}\binom{M_{t}-k-1}{N_{t}-1}\left(\left(1-q_{i}\right)\left(1-\preport\right)\right)^{M_{t}}.\nonumber \\
	= & \tilde{C}\eta_{i}^{k+N_{t}}\binom{k+N_{t}+r_{t}-1}{r_{t}-1}\hypG{N_{t}}{k+N_{t}+r_{t}}{k+N_{t}+1}{\eta_{i}}
\end{align}
where $\eta_{i}=\left(1-q_{i}\right)\left(1-\preport\right)$, $\bar{C}$
and $\tilde{C}$ are normalization constants, and $\hypG{\cdot}{\cdot}{\cdot}{\cdot}$
is Gauss's hypergeometric function. To derive the normalization constant
$\tilde{C}$, we sum over all $k\in\N$:
\begin{align}
	\tilde{C} & =\smft{k=0}{\infty}\smft{M_{t}=N_{t}+k}{\infty}\binom{M_{t}+r_{t}-1}{r_{t}-1}\binom{M_{t}-k-1}{N_{t}-1}\eta_{i}^{M_{t}}\nonumber \\
	& =\smft{M_{t}=N_{t}}{\infty}\binom{M_{t}+r_{t}-1}{r_{t}-1}\eta_{i}^{M_{t}}\smft{k=0}{M_{t}-N_{t}}\binom{M_{t}-k-1}{N_{t}-1}\nonumber \\
	& =\smft{M_{t}=N_{t}}{\infty}\binom{M_{t}+r_{t}-1}{r_{t}-1}\eta_{i}^{M_{t}}{M_{t} \choose N_{t}}\nonumber \\
	& =\binom{N_{t}+r_{t}-1}{r_{t}-1}\frac{\eta_{i}^{N_{t}}}{\left(1-\eta_{i}\right)^{N_{t}+r_{t}}}.
\end{align}
Thus,
\begin{multline}
	\pr{k\text{ unrecorded trips on day }t\,|\,\text{\ensuremath{N_{t}} trips recorded on day }t}\\
	=\frac{\binom{N_{t}+k+r_{t}-1}{r_{t}-1}}{\binom{N_{t}+r_{t}-1}{r_{t}-1}}\eta_{i}^{k}\left(1-\eta_{i}\right)^{N_{t}+r_{t}}\hypG{N_{t}}{k+N_{t}+r_{t}}{k+N_{t}+1}{\eta_{i}}.
\end{multline}

With this result, we can proceed deriving the joint distribution of
unrecorded intermediate trips between two recorded trips. Let $f_{\mt s}\ap{k;t,N_{t}}$
be the probability mass function for the number of unrecorded trips
in one ``section'' of day $t$ given that $N_{t}$ trips were recorded
on the day. Let furthermore $f_{\mt{\mt d}}\ap{k;i,\mathcal{T}}$
be the probability mass function of the random variable modelling
the number of unrecorded trips during the set $\mathcal{T}$ of complete
days. Let $\mathcal{T}_{t_{l}t_{l+1}}$ be the set of days between
$t_{l}$ and $t_{l+1}$. Then the probability mass function of unrecorded
intermediate trips is
\begin{equation}
	f_{\mt{\mt{intm}}}\ap{k;i,t_{l},t_{l+1},N_{t_{l}},N_{t_{l+1}}}=\begin{cases}
		\enskip f_{\mt s}\ap{k;i,t_{l},N_{t_{l}}} & \text{if \ensuremath{t_{l}=t_{l+1}}},\\
		\ap{f_{\mt s}\ap{\tilde{k};i,t_{l},N_{t_{l+1}}}*f_{\mt s}\ap{\tilde{k};i,t_{l+1},N_{t_{l+1}}}}\ap k & \text{if \ensuremath{t_{l+1}-t_{l}=1}}\\
		\ap{f_{\mt s}\ap{\tilde{k};i,t_{l},N_{t_{l+1}}}*f_{\mt s}\ap{\tilde{k};i,t_{l+1},N_{t_{l+1}}}*f_{\mt{\mt d}}\ap{\tilde{k};i,\mathcal{T}_{t_{l}t_{l+1}}}}\ap k & \text{else,}
	\end{cases},
\end{equation}
where ``$*$'' denotes the discrete convolution in $\tilde{k}$. 

To determine the likelihood for an observed sequence of trips, we
need to sum over all possible numbers of unrecorded trips. That is,
we desire to determine 
\begin{equation}
	p_{ij_{l}j_{l+1}|\mathcal{R},t_{l},t_{l+1},N_{t_{l}},N_{t_{l+1}}}=\smft{k=0}{\infty}f_{\mt{\mt{intm}}}\ap{k;i,t_{l},t_{l+1},N_{t_{l}},N_{t_{l+1}}}p_{ij_{l}j_{l+1}|\mathcal{R},k+1}.\label{eq:p-trip-pair-series}
\end{equation}
Note that the argument $k$ of $f_{\mt{\mt{intm}}}$ refers to the
number of intermediate unrecorded trips, whereas the corresponding
subscript of $p_{ij_{l}j_{l+1}|\mathcal{R},k+1}$ refers to the total
number of trips (i.e. all unrecorded trips and one recorded trip). 

To compute (\ref{eq:p-trip-pair-series}) efficiently, we write 
\begin{equation}
	p_{ij_{l}j_{l+1}|\mathcal{R},k+1}=\omega_{ij_{l+1}\mathcal{R}}+\psi_{ij_{l}j_{l+1}\mathcal{R}}\psame^{k+1}
\end{equation}
with 
\begin{align}
	\omega_{ij_{l+1}\mathcal{R}} & =\pall p_{ij_{l+1}}+\I{\mathcal{R}}{j_{l+1}}\pregion p_{ij_{l+1}|\mathcal{R}}\\
	\psi_{ij_{l}j_{l+1}\mathcal{R}} & =\left(1-\delta_{1l}\right)\left(\pall p_{ij_{l+1}}+\I{\mathcal{R}}{j_{l+1}}\pregion p_{ij_{l+1}|\mathcal{R}}\right)+\delta_{j_{l}j_{l+1}}.
\end{align}
Then,

\begin{align*}
	p_{ij_{l}j_{l+1}|\mathcal{R},t_{l},t_{l+1},N_{t_{l}},N_{t_{l+1}}} & =\smft{k=0}{\infty}f_{\mt{\mt{intm}}}\ap{k;i,t_{l},t_{l+1},N_{t_{l}},N_{t_{l+1}}}\left(\omega_{ij_{l+1}\mathcal{R}}+\psi_{ij_{l}j_{l+1}\mathcal{R}}\psame^{k+1}\right)\\
	& =\omega_{ij_{l+1}\mathcal{R}}+\psi_{ij_{l}j_{l+1}\mathcal{R}}\Upsilon_{t_{l}t_{l+1}N_{t_{l}}N_{t_{l+1}}}\Lambda_{t_{l}t_{l+1}N_{t_{l}}N_{t_{l+1}}}
\end{align*}
with 
\begin{flalign}
	\Upsilon_{it_{l}t_{l+1}N_{t_{l}}N_{t_{l+1}}} & =\begin{cases}
		\tilde{g}_{\mt s}\ap{i,t_{l},N_{t_{l}}} & \text{if \ensuremath{t_{l}=t_{l+1}}},\\
		\tilde{g}_{\mt s}\ap{i,t_{l},N_{t_{l}}}\tilde{g}_{\mt s}\ap{i,t_{l+1},N_{t_{l+1}}} & \text{if \ensuremath{t_{l+1}-t_{l}=1}}\\
		\left(1-\eta_{i}\right)^{\bar{r}_{\mathcal{T}_{t_{l}t_{l+1}}}}\tilde{g}_{\mt s}\ap{i,t_{l},N_{t_{l}}}\tilde{g}_{\mt s}\ap{i,t_{l+1},N_{t_{l+1}}} & \text{else,}
	\end{cases}\\
	\Lambda_{it_{l}t_{l+1}N_{t_{l}}N_{t_{l+1}}} & =\begin{cases}
		\smft{k=0}{\infty}\tilde{f}_{\mt s}\ap{\tilde{k};i,t_{l},N_{t_{l}}}\xi_{0}^{k+1} & \text{if \ensuremath{t_{l}=t_{l+1}}},\\
		\smft{k=0}{\infty}\left(\tilde{f}_{\mt s}\ap{\tilde{k};i,t_{l},N_{t_{l}}}*\tilde{f}_{\mt s}\ap{\tilde{k};i,t_{l+1},N_{t_{l+1}}}\right)\left(k\right)\xi_{0}^{k+1} & \text{if \ensuremath{t_{l+1}-t_{l}=1}}\\
		\smft{k=0}{\infty}\left(\tilde{f}_{\mt s}\ap{\tilde{k};i,t_{l},N_{t_{l}}}*\tilde{f}_{\mt s}\ap{\tilde{k};i,t_{l+1},N_{t_{l+1}}}*\binom{\tilde{k}+\bar{r}_{\mathcal{T}_{t_{l}t_{l+1}}}-1}{\bar{r}_{\mathcal{T}_{t_{l}t_{l+1}}}-1}\right)\left(k\right)\xi_{0}^{k+1} & \text{else,}
	\end{cases}
\end{flalign}
and 
\begin{align}
	\tilde{g}_{\mt s}\ap{i,t,N_{t}} & =\left(1-\eta_{i}\right)^{N_{t}+r_{t}}\left(\binom{N_{t}+r_{t}-1}{r_{t}-1}\right)^{-1}\\
	\tilde{f}_{\mt s}\ap{i,k;t,N_{t}} & =\binom{N_{t}+k+r_{t}-1}{r_{t}-1}\hypG{N_{t}}{k+N_{t}+r_{t}}{k+N_{t}+1}{\eta_{i}}\eta_{i}^{k}.
\end{align}

For each combination of origin $i$, day $t$, and recorded number
of trips $N_{t}$ we need to evaluate $\tilde{g}_{\mt s}\ap{i,t,N_{t}}$
and $\tilde{f}_{\mt s}\ap{k;i,t,N_{t}}$ only once. The same applies
to day- and record-number pairs and $\Upsilon_{t_{l}t_{l+1}N_{t_{l}}N_{t_{l+1}}}$
and $\Lambda_{t_{l}t_{l+1}N_{t_{l}}N_{t_{l+1}}}$. This can help to
speed up computations, if multiple anglers have recorded the same
numbers of trips (typically one) on the same days. As we are not aware
of a closed-form expression for the series $\Lambda_{it_{1}t_{2}N_{1}N_{2}}$,
we cannot compute it exactly. However, if we only consider a moderate
number of terms, the truncation error is very small. We considered
the first $200$ terms of the sum. 

With the above result, we can proceed deriving the joint likelihood
for a sequence of observations. First, recall the likelihood for the
temporal sequence of trips derived in equation (\ref{eq:time-sequence-pmf}).
Second, note that since we do not know the anglers' preferred fishing
region, we need to sum over all region candidates $\mathcal{R}$ to
determine the correct joint probability for multiple observations.
Furthermore, we need to consider that not all anglers are using the
app. The probability that a sequence $j_{1},\dots,j_{m}$ of trips
is recorded by the app for an angler from locality $i$ is then 
\begin{align}
	p_{ij_{1}\dots j_{m}} & =\papp\tilde{q}_{i}^{T/\alpha}\pro{t\,:\,k_{t}\neq0}\binom{k_{t}+r_{t}-1}{r_{t}-1}\ap{1-\tilde{q}_{i}}^{k_{t}}\nonumber \\
	& \qquad\smo{\mathcal{R}}p_{i\mathcal{R}}\prft{l=1}m\left(\omega_{ij_{l+1}\mathcal{R}}+\psi_{ij_{l}j_{l+1}\mathcal{R}}\Upsilon_{t_{l}t_{l+1}N_{t_{l}}N_{t_{l+1}}}\varLambda_{t_{l}t_{l+1}N_{t_{l}}N_{t_{l+1}}}\right)\label{eq:likelihood-single-angler}
\end{align}
with $\tilde{q}_{i}=1/\left(1+\bar{\alpha}\preport\mu_{i}\right)$. 

It is inefficient to sum over all potential regions of preference
as suggested in equation (\ref{eq:likelihood-single-angler}), but
this is not necessary to compute the likelihood. Note that $\omega_{ij_{l+1}\mathcal{R}}$
and $\psi_{ij_{l}j_{l+1}\mathcal{R}}$ are independent of the region
$\mathcal{R}$ if the destination $j_{l+1}$ is not in $\mathcal{R}$.
Let $\mathfrak{R}_{j}=\left\{ \mathcal{R}\,|\,j\in\mathcal{R}\right\} $
be the set of all regions that contain the destination $j$. Then,
\begin{align}
	& \smo{\mathcal{R}}p_{i\mathcal{R}}\prft{l=1}m\left(\omega_{ij_{l+1}\mathcal{R}}+\psi_{ij_{l}j_{l+1}\mathcal{R}}\Upsilon_{t_{l}t_{l+1}N_{t_{l}}N_{t_{l+1}}}\varLambda_{t_{l}t_{l+1}N_{t_{l}}N_{t_{l+1}}}\right)\nonumber \\
	= & \smo{\mathcal{R}\in\uno l\mathfrak{R}_{l}}p_{i\mathcal{R}}\prft{l=1}m\left(\omega_{ij_{l+1}\mathcal{R}}+\psi_{ij_{l}j_{l+1}\mathcal{R}}\Upsilon_{t_{l}t_{l+1}N_{t_{l}}N_{t_{l+1}}}\varLambda_{t_{l}t_{l+1}N_{t_{l}}N_{t_{l+1}}}\right)\nonumber \\
	& \qquad+\smo{\mathcal{R}\notin\uno l\mathfrak{R}_{l}}p_{i\mathcal{R}}\prft{l=1}m\left(\omega_{ij_{l+1}\emptyset}+\psi_{ij_{l}j_{l+1}\emptyset}\Upsilon_{t_{l}t_{l+1}N_{t_{l}}N_{t_{l+1}}}\varLambda_{t_{l}t_{l+1}N_{t_{l}}N_{t_{l+1}}}\right)\nonumber \\
	= & \smo{\mathcal{R}\in\uno l\mathfrak{R}_{l}}p_{i\mathcal{R}}\prft{l=1}m\left(\omega_{ij_{l+1}\mathcal{R}}+\psi_{ij_{l}j_{l+1}\mathcal{R}}\Upsilon_{t_{l}t_{l+1}N_{t_{l}}N_{t_{l+1}}}\varLambda_{t_{l}t_{l+1}N_{t_{l}}N_{t_{l+1}}}\right)\nonumber \\
	& \qquad+\left(1-\smo{\mathcal{R}\in\uno l\mathfrak{R}_{l}}p_{i\mathcal{R}}\right)\prft{l=1}m\left(\omega_{ij_{l+1}\emptyset}+\psi_{ij_{l}j_{l+1}\emptyset}\Upsilon_{t_{l}t_{l+1}N_{t_{l}}N_{t_{l+1}}}\varLambda_{t_{l}t_{l+1}N_{t_{l}}N_{t_{l+1}}}\right)\label{eq:RegionSum}
\end{align}
with 
\begin{align}
	\omega_{ij_{l+1}\emptyset} & =p_{ij_{l+1}}\pall\\
	\psi_{ij_{l}j_{l+1}\emptyset} & =\delta_{j_{l}j_{l+1}}-\left(1-\delta_{1l}\right)p_{ij_{l+1}}\pall.
\end{align}
Since $\uno l\mathfrak{R}_{l}$ is typically small, the sum (\ref{eq:RegionSum})
can be computed efficiently.

If no trip was observed for an angler, this may have three causes:
(1) the angler may not use the app, (2) the angler may have decided
not to record any of their trips, and (3) the angler may not have
made any trips. Hence, the probability to record no trip by an angler
from origin $i$ is 
\begin{equation}
	p_{i\emptyset}=\left(1-\papp\right)+\papp\tilde{q}_{i}^{T/\alpha}.
\end{equation}
The overall joint likelihood function of our model is the product
of the probabilities $p_{ij_{1}\dots j_{m}}$ or $p_{i\emptyset}$
for all anglers. 

The likelihood function derived above does not include the parameter
$\papp$, since anglers who do not use the app would not be in our
dataset. In turn, the likelihood function includes the parameter $\mu_{i}$,
which could only be estimated up to a proportionality constant in
the previous step. Nonetheless, both $\papp$ and $\mu_{i}$ can be
estimated via our model. Recall from Appendix \ref{subsec:APX-Likelihood-Gravity}
that we are able to estimated the constant $C=\papp\preport c$, where
$c$ is the proportionality constant of $\mu_{i}$. The constant $\preport$
occurs in the likelihood function described and may thus be estimated.
Furthermore, the likelihood function includes $\mu_{i}$, which can
be written as $\mu_{i}=c\tilde{\mu}_{i}$, where $\tilde{\mu}_{i}$
has is estimated via the gravity model (see Appendix \ref{subsec:APX-Likelihood-Gravity}).
Hence, $c$ appears in the likelihood function as a free parameter
and may thus be estimated. With the estimates of $\preport$, $c$,
and $C$, however, the fraction of app users $\papp$ can be computed
directly. That is, all parameters of the model may be estimable. 

\section{Details about the procedures applied to optimize likelihood and AIC\label{APX-optimization}}

In this appendix, we provide details about the procedures we applied
to maximize the likelihood functions and to select the model with
the minimal AIC value. 

\subsection{Likelihood maximization}

To maximize the likelihood, we used a number of different optimizations
algorithms with different strengths that we applied consecutively.
For all submodels, we started with a meta-heuristic global search
algorithm in a generous subspace of the parameter space. Specifically,
we used the ``differential evolution'' algorithm by \citet{storn_differential_1997}.
As this algorithm converges only slowly, we terminated it after a
moderate number of iterations. Then we proceeded with the L-BFGS-G
algorithm \citep{byrd_limited_1995}, which was robust and efficient
even when differential evolution algorithm returned a solution far
from the optimum. With the result as initial condition, we applied
sequential least squares programming \citep{kraft_software_1988},
which converged quickly to a neighbourhood of the respective likelihood
maximum. Finally, we refined the result with a trust-region Newton-Raphson
method \citep{nocedal_trust-region_2006}, which is particularly efficient
in the neighbourhood of the maximum. Only when we optimized the model
for the day suitability, we skipped the L-BFGS-G and the sequential
least squares programming step, because the model had a simple likelihood
function and only few parameters, which made it easy to fit the model.
For all applied algorithms, we used the implementations from the scientific
computing library Scipy \citep{jones_scipy:_2001}. 

Whenever needed, we determined derivatives of the log-likelihood functions
via automatic differentiation. To that end, we used the python package
``autograd''. However, some special functions in the log-likelihood
function of the submodel for the choice parameters (see Appendix \ref{subsec:APX-Likelihood-Choice-Parameters})
were not included in the automatic differentiation package. Hence,
we approximated derivatives numerically, using the package ``numdifftools''. 

To avoid numerical issues arising when optimizers step outside of
the domain of the likelihood function, we mapped parameters to the
domain as follows: when a parameter $x$ was constrained to be positive,
it was expressed as 
\begin{equation}
	x=\begin{cases}
		\exp\tilde{x} & \text{if \ensuremath{\tilde{x}<0}}\\
		\tilde{x}+1 & \text{else,}
	\end{cases}
\end{equation}
and if it was constrained to the interval $\left[0,1\right]$, it
was expressed as 
\begin{equation}
	x=\frac{1}{\pi}\arctan\tilde{x}+\frac{1}{2}.
\end{equation}
Then, the transformed unconstrained parameter $\tilde{x}$ was optimized,
respectively. 

Most parameters we considered had a ``natural'' domain. For example,
all parameters representing probabilities are constrained to the interval
$\left[0,1\right]$. However, the domain of the parameters of the
gravity model used to estimate the angler activeness and the destination
choice probabilities may be less clear. We constrained the ``factor''
parameters $\beta_{k}$ (see equations (9) and (10)
in the main text) to positive values to avoid the possibility to obtain
negative trip rates. The ``exponent'' parameters $\gamma_{k}$ were
constrained to positive values if and only if the respective covariate
included zeros or very small values. This was the case for all covariates
referring to the subbasins.

\subsection{AIC optimization\label{subsec:APX-AIC-optimization}}

Our gravity model for the angler activeness and the destination choice
probabilities can involve a large number of parameters. Considering
all possible combinations of parameters to select the model with the
minimal AIC would not be feasible. However, the search can be sped
with a branch and bound type search, which we describe below.

For a set $\mathcal{P}$ of parameters let $\ell_{\mathcal{P}}^{*}$
be the maximal log-likelihood that can be achieved if these parameters
are included in the model and $\mt{AIC}_{\mathcal{P}}$ the corresponding
AIC value. Suppose we have already determined the AIC value of some
models, where $\mt{AIC}^{*}$ is the lowest AIC value we have achieved
so far. Since we are considering nested models, exclusion of a parameter
can never increase the likelihood. Hence, for any subset $\mathcal{P}'\subset\mathcal{P}$
it is $\ell_{\mathcal{P}'}^{*}\leq\ell_{\mathcal{P}}^{*}$. By definition
of the AIC, it is 
\begin{equation}
	\mt{AIC}_{\mathcal{P}}=2\left(\left|\mathcal{P}\right|-\ell_{\mathcal{P}}^{*}\right),\label{eq:AIC-def}
\end{equation}
where $\left|\mathcal{P}\right|$ denotes the number of elements in
the set $\mathcal{P}$. 

Let $\mathfrak{P}$ be the set of all parameter sets that we have
considered so far. Suppose $\mathcal{P}'\notin\mathfrak{P}$ is a
parameter set that leads to a better AIC value then the currently
best one: $\mt{AIC}_{\mathcal{P}}<\mt{AIC}^{*}$. For any $\mathcal{P}\in\mathfrak{P}$
with $\mathcal{P}'\subset\mathcal{P}$ it holds then
\begin{equation}
	\mt{AIC}^{*}>2\left(\left|\mathcal{P}'\right|-\ell_{\mathcal{P}'}^{*}\right)\geq2\left(\left|\mathcal{P}'\right|-\ell_{\mathcal{P}}^{*}\right).
\end{equation}
Thus, 
\begin{equation}
	\left|\mathcal{P}'\right|<\maxo{\mathcal{P}\in\mathfrak{P}\,:\,\mathcal{P}\supset\mathcal{P}'}\ell_{\mathcal{P}}^{*}+\frac{1}{2}\mt{AIC}^{*},\label{eq:AIC-condition}
\end{equation}
and we can ignore all parameter combinations that do not satisfy this
condition. 

To minimize the AIC value, we start with the full model (including
all possible parameters). Then, we consider all models for which one
parameter is left out, then all models for which two parameters are
left out end so on. In doing so, we skip all models that do not satisfy
condition (\ref{eq:AIC-condition}) with the best currently found
AIC value. We terminate the search when no admissible parameter set
is left.

In the worst case, the search may include an extremely large number
of models that cannot be considered in reasonable time. In practice,
however, a certain set of parameters improves the model fit by a lot,
so that only combinations in which the remaining parameters are left
out need to be considered. Hence, a complete search is often possible
in practice. 

Since it would require an exponential number of computation steps
to explicitly check for each possible parameter combination whether
condition (\ref{eq:AIC-condition}) is satisfied, this is not a viable
option in practice. Instead, we store all already considered parameter
sets in a trie data structure \citep{aoe_efficient_1989}, which enables
us to efficiently generate only those parameter combinations that
satisfy condition (\ref{eq:AIC-condition}). That way, a complete
search remains feasible even if the a large number of parameters are
considered.

\section{Computing the expected mean yearly fishing days per angler \label{sec:APX-days-out}}

Below, we show how the mean expected number of days that anglers in
Alberta go fishing can be computed based on our model. Consider an
angler from locality $i$. The number of their trips on a given day
$t$ is negative binomially distributed with mean $\mu_{i}\tau_{t}$
and dispersion parameter $\tau_{t}/\alpha$. Hence, the probability
that they made at least one trip on day $t$ is
\begin{align}
	\pr{\text{\#trips on day }t\geq1} & =1-\pr{\text{\#trips on day }t=0}\nonumber \\
	& =1-\left(\frac{1}{1+\alpha\mu_{i}}\right)^{\frac{\tau_{t}}{\alpha}}.
\end{align}
Let $\mathcal{T}$ be the set of days in the year of interest. Then
the expected mean number of yearly fishing days is given by 
\begin{equation}
	\frac{\smo i\smo{t\in\mathcal{T}}n_{i}\left(1-\left(\frac{1}{1+\alpha\mu_{i}}\right)^{\frac{\tau_{t}}{\alpha}}\right)}{\smo in_{i}}.
\end{equation}

\section{The effect of an overestimated dispersion parameter on model estimates\label{sec:APX-overdispersion-bias}}

In this appendix, we show how overestimating the dispersion parameter
affects the estimated mean traffic counts. The dispersion parameter
may be overestimated if no data replicates are available to fit the
model (e.g. from a longitudinal study) and errors due to a misspecified
model are mistakenly attributed to stochasticity. \citet{piegorsch_maximum_1990}
noted that the log-likelihood function for $n$ independent samples
$y_{i}$ from a negative binomial distribution with dispersion parameter
$\alpha$ and mean $\mu$ can be written as 
\begin{equation}
	\ell\ap{\alpha,\mu}=\frac{1}{n}\smft{i=1}n\left(\smft{\nu=0}{y_{i}-1}\ln\ap{1+\alpha\mu}+y_{i}\ln\mu-\left(y_{i}+\alpha^{-1}\right)\ln\ap{1+\alpha\mu}\right),
\end{equation}
with partial derivative 
\begin{alignat}{1}
	\pder[\mu]{}\ell\ap{\alpha,\mu} & =\frac{1}{n}\smft{i=1}n\left(\frac{y_{i}}{\mu}-\frac{1+\alpha y_{i}}{1+\alpha\mu}\right)\nonumber \\
	& =\frac{1}{n}\smft{i=1}n\frac{y_{i}-\mu}{\mu\left(1+\alpha\mu\right)}.
\end{alignat}

Suppose that each observation $y_{i}$ has been drawn from a distribution
with different mean $\mu_{i}$, which is computed based on a covariate
vector $c_{i}$ according to a model with parameter vector $\th$:
\begin{equation}
	\mu_{i}=f\ap{\th;c_{i}}.
\end{equation}
When the likelihood is maximized with respect to $\th$, the partial
derivatives with respect to the components $\th_{i}$ must be $0$:
\begin{align}
	0 & =\pder[\th_{j}]{}\ell\ap{\alpha,\mu_{1},\dots,\mu_{n}}\nonumber \\
	& =\smft{i=1}n\pder[\th_{j}]{\mu_{i}}\pder[\mu_{i}]{\ell}\nonumber \\
	& =\frac{1}{n}\smft{i=1}n\pder[\th_{j}]{\mu_{i}}\frac{y_{i}-\mu_{i}}{\mu_{i}\left(1+\alpha\mu_{i}\right)}.\label{eq:score-0}
\end{align}
Clearly, if the model fits the data perfectly, then
\begin{equation}
	\mu_{i}=y_{i}\label{eq:likelihood-optim}
\end{equation}
solves (\ref{eq:score-0}) regardless of the value of $\alpha$. However,
if the model is slightly misspecified or (\ref{eq:likelihood-optim})
cannot be achieved due to the stochastic nature of the observations
$y_{i}$, the dispersion parameter has an effect on the estimates
of the mean values $\mu_{i}$. With small overdispersion, $\alpha\rightarrow0$,
equation (\ref{eq:score-0}) becomes approximately
\begin{equation}
	0=\smft{i=1}n\pder[\th_{j}]{\mu_{i}}\frac{y_{i}-\mu_{i}}{\mu_{i}},
\end{equation}
whereas we obtain with large overdispersion, $\alpha\rightarrow\infty$,
\begin{equation}
	0=\smft{i=1}n\pder[\th_{j}]{\mu_{i}}\frac{y_{i}-\mu_{i}}{\mu_{i}^{2}}.
\end{equation}
That is, when fitting the model, values with small predicted mean
are weighed higher if the overdispersion is large. Consequently, if
the dispersion parameter $\alpha$ is overestimated, estimates from
the fitted model will match small values more accurately than large
values. When estimating traffic flows, this is of two-fold concern:
(1) Often, large traffic flows are of higher interest than small ones.
Hence, it is of disadvantage when the model fits traffic flows with
small mean better than traffic flows with high mean. (2) Since trips
are rare events for weak traffic flows, the corresponding trip counts
are subject to significant error. Unless a very large dataset is available,
putting a disproportionate weight on these highly stochastic observations
may lead to overfitting. 

\includegraphics[scale=0.01]{dummy.pdf}

\bibliographystyle{mee}